\pdfoutput=1
\documentclass[twocolumn]{article}

\usepackage{algorithm}
\usepackage{algorithmic}
\usepackage{graphicx}
\usepackage{subfig}
\date{}

\usepackage{amssymb}
\usepackage{amsmath}
\usepackage[latin2]{inputenc}
\usepackage{verbatim}
\usepackage{framed}
\usepackage{url}

\title{Cloud Workload Prediction based on Workflow Execution Time Discrepancies}
\author{Gabor Kecskemeti\footnote{Department of Computer Science, Liverpool John Moores University; email: g.kecskemeti@ljmu.ac.uk} \and Zsolt Nemeth\footnote{Laboratory of Parallel and Distributed Systems, MTA SZTAKI} \and Attila Kertesz\footnote{Software Engineering Department, University of Szeged} \and Rajiv Ranjan\footnote{School of Computing, Newcastle University}}
\begin{document}

\maketitle
\begin{abstract}
Infrastructure as a service clouds hide the complexity of maintaining the physical infrastructure with a slight disadvantage: they also hide their internal working details. Should users need knowledge about these details e.g., to increase the reliability or performance of their applications, they would need solutions to detect behavioural changes in the underlying system. Existing runtime solutions for such purposes offer limited capabilities as they are mostly restricted to revealing weekly or yearly behavioural periodicity in the infrastructure. This article proposes a technique for predicting generic background workload by means of simulations that are capable of providing additional knowledge of the underlying private cloud systems in order to support activities like cloud orchestration or workflow enactment. Our technique uses long-running scientific workflows and their behaviour discrepancies and tries to replicate these in a simulated cloud with known (trace-based) workloads. We argue that the better we can mimic the current discrepancies the better we can tell expected workloads in the near future on the real life cloud. We evaluated the proposed prediction approach with a biochemical application on both real and simulated cloud infrastructures. The proposed algorithm has shown to produce significantly ($\sim$20\%) better workload predictions for the future of simulated clouds than random workload selection.
\end{abstract}

\section{Introduction}

Infrastructure as a Service (IaaS) clouds became the foundations of compute/data intensive applications \cite{cloudIaaS}. They provide computational and storage resources in an on demand manner. The key mechanism of IaaS is virtualisation that abstracts resource access mechanisms with the help of Virtual Machines (VM) allowing their users to securely share physical resources. While IaaS clouds offer some means to control a virtual ensemble of resources (so called virtual infrastructures), they inherently provide no means for precise insight into the state, load, performance of their resources, thus the physical layer is completely hidden. Due to the multi-tenant environment of clouds, application performance may be significantly affected by other, (from the point of view of a particular user) unknown and invisible processes, the so-called background workload. Albeit Service Level Agreements (SLA) define the expected specifics and various Quality of Service (QoS) methods are aimed at their fulfilment, yet they can provide a very broad range of performance characteristics only \cite{Kritikos13}. 

This article studies performance issues related to the -- unknown -- background load and proposes a  methodology for its estimation. We envision a scenario where modifications in the virtual infrastructure are necessary at runtime and to make the right decisions and take actions the background load cannot be omitted. As follows, we made two assumptions: (i) the application runs long enough so that the time taken by a potential virtual infrastructure re-arrangement is negligible and (ii) the application is executed repeatedly over a period of time. Both these assumptions are valid for a considerably large class of cloud based applications. Scientific workflows are especially good candidates for exemplifying this class as they are executed in numerous instances by large communities over various resources~\cite{szabo}. During execution, jobs of a workflow are mapped onto various resources e.g.,  a parallel computer, a cluster, a grid, a virtual infrastructure on a cloud, etc. Efficient execution of workflows requires a precise scheduling of tasks and resources which furthermore, requires both timely information on the resources and the ability to control them. Thus we have chosen scientific workflows as a subject and evaluation example of our method.

The recurring nature of workflows enables the extraction of performance data and also successive adaptation, refinement and optimisation leading to dynamic workflow enactment. The main motivation for our work stems from the assumption that by extracting information from {\em past} workflow executions, one could {\em identify current} and {\em predict future} background workloads of the resources allocated for the workflow. The result of this prediction subsequently enables to steer current and future cloud usage accordingly, including the option of resource re-arrangement if indicated. The idea is centred around a set of past load patterns (a database of historic traces). When a workflow is being enacted, some of its jobs have already been executed and some others are waiting for execution. Our workload prediction aims at finding historic traces, that likely resemble the background of workload behind the currently running workflow. Hence, future tasks (even those that are completely independent from the workflow that was used for the prediction) are enacted taking into consideration the recent background load estimations.

The main contributions of this article are: (i) the concept of a private-cloud level load prediction method based on the combination of historic traces, aimed at improving execution quality (ii) an algorithm for realising the load prediction at runtime so that performance constraints are observed, and (iii) an evaluation of this approach using a biochemical application with simulations using historic traces from a widely used archive.

The remainder of this article is as follows: Section~\ref{sec:related} presents related work, then Section~\ref{sec:background} introduces the basic terminology and assumptions of our research. Section~\ref{sec:core} introduces our new algorithm. Section~\ref{sec:eval} presents its evaluation with a biochemical application. Finally, the contributions are summarised in Section~\ref{sec:conclusion}.

\section{Related Work}\label{sec:related}

In this article, we examine past traces of certain workflows, and predict the expected background load of the clouds behind current workflow instances. Our technique fits in the analyse phase of autonomous control loops (like monitor-analyse-plan-execute~\cite{autonomic}). Similarly, Maurer et al. \cite{maurer} investigated adaptive resource configuration from a SLA/QoS point of view using such a loop. In their work, actions to fine tune virtual machine (VM) performance are categorised hierarchically as so called escalation levels. Generally, our work addresses a similar problem (our scope is on the background workload level instead of infrastructure and resource management) with a different grained action set for the plan-execute steps of the autonomous loop.

Concerning workload modelling, Khan et al. \cite{khan12} used data traces obtained from a data centre to characterise and predict workload on VMs. Their goal was to explore cross-VM workload correlations, and predict workload changes due to dependencies among applications running in different VMs -- while we approach the load prediction from the workflow enactment point of view.

Li et al. \cite{li11} developed CloudProphet to predict legacy application performance in clouds. This tool is able to trace the workload of an application running locally, and to replay the same workload in the cloud for further investigations and prediction. In contrast, our work presents a technique to identify load characteristics independent from the workflow ran on cloud resources.

Fard et al. \cite{Fard16} also identified performance uncertainties of multi-tenant virtual machine instances over time in Cloud environments. They proposed a model called R-MOHEFT that considered uncertainty intervals for workflow activity processing times. They developed a three-objective (i.e., makespan, monetary cost, and robustness) optimisation for Cloud scheduling in a commercial setting. In contrast to this approach our goal is to identify patterns in earlier workloads to overcome the uncertainty, and apply simulations to predict future background load of the infrastructure.

Calheiros et al. \cite{calheiros15} offers cloud workload prediction based on autoregressive integrated moving average. They argue that proactive dynamic provisioning of resources could achieve good quality of service. Their model's accuracy is evaluated by predicting future workloads of real request traces to web servers. Additionally, Magalhaes et al. \cite{magalhaes} developed a workload model for the CloudSim simulator using generalised extreme value/lambda distributions. This model captures user behavioural patterns and supports the simulation of resource utilisation in clouds. They argue that user behaviour must be considered in workload modelling to reflect realistic conditions. Our approach share this view: we apply a runtime behaviour analysis to find a workflow enactment plan that best matches the infrastructure load including user activities. 

Caron et al. \cite{caron} used workload prediction based on identifying similar past occurrences of the current short-term workload history for efficient resource scaling. This approach is the closest to ours (albeit, we have a different focus support for on-line decision making in scientific workflow enactors etc.), as it uses real-world traces from clouds and grids. They examine historic data to identify similar usage patterns to a current window of records, and their algorithm predicts the system usage by extrapolating beyond the identified patterns. In contrast, our work's specific focus on scientific workflows allows the analysis and prediction of recently observed execution time discrepancies, by introducing simulations to the prediction and validation phases. 

Pietri et al. \cite{pietri} designed a prediction model for the execution time of scientific workflows in clouds. They map the structure of a workflow to a model based on data dependencies between its nodes to calculate an estimated makespan. Though the goal of this paper, i.e. to determine the amount of resources to be provisioned for better workflow execution based on the proposed prediction method is the same in our article, we rely on the runtime workflow behaviour instead of its structure. This means we aim to predict the background load instead of the execution time of a workflow.  

\section{Background}
\label{sec:background}

An {\em enactment plan} describes the jobs of a scientific workflow, their schedule to resources and it is processed by a workflow enactor that does the necessary assignments between jobs and resources. If a workflow enactor is capable to handle dynamic environments~\cite{Caballer}, such as clouds, the resources form a virtual infrastructure (crafted to serve specific jobs). In our vision, the enactment plan also lists the projected execution time of each job in the workflow. Workflow enactors are expected to base the projected execution time on historic executions to represent their expectations wrt. the job execution speed. This enactment extension allows the workflow enactor to offer background knowledge on the behaviour  past runs of the workflow that combined the use of various distinct inputs and resource characteristics. As a result, during the runtime of the workflow, infrastructure provisioning issues could be pinpointed by observing  deviations from the projected execution time in the enactment plan.

The virtual infrastructures created by the enactor are often hosted at IaaS cloud providers  that tend to feature multi-tenancy and under provisioning for optimal costs and resource utilisation. These practices, especially under provisioning, could potentially hinder the virtual infrastructure's performance (and thus the execution times of jobs allocated to them). In accordance with the first phase ({\em monitor}) of autonomous control loops, to maintain the quality and to meet the SLAs set out for the virtual infrastructure in the enactment plan, the workflow enactor or a third party service continuously monitors the behaviour of the applications/services/workflows running on the virtual infrastructure. In case of deviations, actions in the management of the virtual infrastructure should take place, such as adding or removing new computing/storage components, to minimize fluctuations in the \emph{quality of execution} (note: these reactive actions are out of scope of this article). We assume sufficiently small, likely private, cloud infrastructures where the workflow instances could experience significant enough portion of the whole infrastructure allowing the exploitation of the identified deviations for prediction purposes. 

We represent workflows $W \in \mathcal{W}$ (where $\mathcal{W}$ is the set of all possible abstract workflows) as an ordered set of jobs: $W=\{j_1\dots j_N\}$, where the total number of jobs in the workflow is $N\in\mathbb{N}$.  The job order is set by their projected completion time on the virtual infrastructure whereas the job inter-relations (dependencies) are kept in the domain of the workflow enactors. The projected execution time of the a job ($j_x\in W$) is  $r_{ex}(j_x)$~--~where $r_{ex}:W\to\mathbb{R^+}$. We expect the enactor to calculate the projected execution times based on its background knowledge about thousands of past runs.

We refer to a workflow instance (i.e., a particular execution of the abstract workflow $W$) with the touple: $[W,t]: \mathcal{W} \times \mathcal{T}$ -- i.e., the workflow and the start time ($t$ and $\mathcal{T}$ depicts the set of all time instances) of its first job $[j_1,t]$. Hence, all instances of $j_x \in W$ are also identified as $[j_x,\mathcal{T}]: j_x \in [W, \mathcal{T}]$. Once the workflow started, the enactor's monitoring facilities will collect the observed execution times for each job instance. We denote these as: $r_{ob}(j_x,t)$ -- where $r_{ob}:W\times \mathcal{T}\to\mathbb{R^+}$. 

Using the acquired data from the enactors and its monitoring facilities, we define the \emph{error function of (partial) workflow execution time} to determine the deviation in execution time of a particular workflow suffered compared to the projected times in the enactment plan. Such function is partial if the evaluated workflow instance is split into two parts: jobs $j_1,... j_k$ already executed whereas $j_{k+1},... j_N$ are not yet complete or waiting for execution; when $k=N$ the workflow instance is done and the error function determines its final execution time error. So in general, the error function of workflow execution time is defined as: $E: \mathcal{W}\times \mathcal{T}\times\mathbb{N}\to\mathbb{R^+}$. 

We require that error functions assign higher error values for workflow instances that deviate more from the projected runtime behaviour set in their enactment plan. These functions should also penalise both positive and negative execution time differences ensuring that the execution strictly follows the plan. The penalties applied here will allow us to detect if the background workload reduces/improves the performance of the workflow instance compared to the enactor's expectations. These penalties are exploited by the later discussed workload prediction technique: it can tell if a particular  workload estimate is not sufficiently close to its actual real life counterpart. For example, when the execution times show improvements -- negative differences -- under a particular workload estimates, then the prediction technique knows it still has a chance to improve its estimate (allowing other, not necessarily long running, applications to better target the expected background workload on the cloud of the workflow).

The penalties are also important from the point of view of the workflow enactor. The enactment plan likely contains projected values resulted from several conflicting service level requirements (regarding the \emph{quality of the execution}). These projected values are carefully selected by the enactor to meet the needs of the workflow's user and follow the capabilities of the used cloud resources. Thus, error functions should indicate if the fulfilment of the projected values set by the enactor are at risk (ie., they should penalize with higher error values even if the observed execution times turned out to be better than originally planned). For example, if we have jobs $j_x$ and $j_y$ where $j_y$ is dependent on the output of $j_x$ and several other factors, these factors could make it impossible for $j_y$ to be ready for execution by a the time a better performing $j_x$ completes. Thus, the enactor could make a plan relaxed for $j_x$ and explicitly ask for its longer job execution time. If the job is executed more rapidly in spite of this request, the penalty of the error function would show there were some unexpected circumstances which made the job faster.

\begin{figure}[tb]
	\centering
	\includegraphics[clip=true,trim=2.2cm 9.9cm 10.8cm 2.8cm,width=0.48\textwidth]{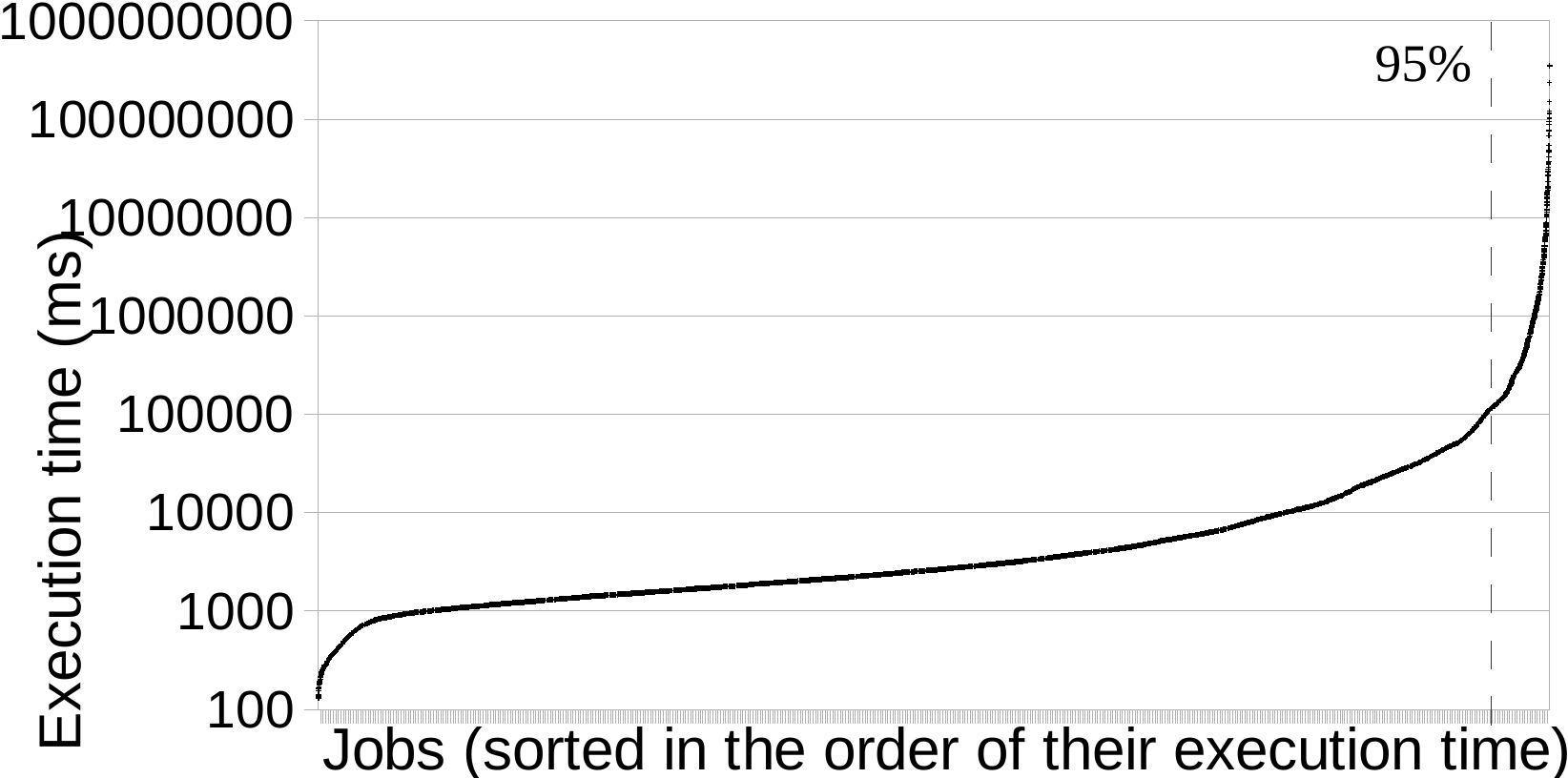}
	\caption{Example distribution of job execution times in a scientific workflow (where 95\% of the execution time is spent on 5 \% of the jobs)\label{FIG-EXPAR}}
\end{figure}

The deviation from the projected execution times (as indicated by the error function) could either be caused by (i) an unforeseen reaction to a specific input, or by (ii) the background load behind the virtual infrastructure of the workflow. In case (i), the input set causes the observed execution times to deviate from the projected ones. Such deviations are rare, because job execution times on a dedicated infrastructure (i.e., only dependent on input values) usually follow a Pareto distribution~\cite{Bacso2013}. Thus, job execution times mostly have small variances, but there could be jobs running several orders of magnitude slower than usual. Figure~\ref{FIG-EXPAR} exemplifies this behaviour with a sample of over 20k jobs ran for various cloud simulation workflows. As it can be observed on the example figure, the long execution times in the slowest 5 \% of the jobs cannot be mistaken for perturbations caused by background load. 

On the other hand (ii), under-provisioning in IaaS clouds can cause significant background load variation yielding observable (but minor) perturbations in job execution times. In this article, we focus on case (ii) only, therefore we must filter observed execution times whether they belong to case (i) or (ii). Consequently, when observing a significant increase in job execution time (i.e., enactor's predicted execution time is a magnitude smaller than what was actually observed), we assume that the particular job belongs to case (i) and we do not apply our technique. However, when we only observe minor deviations from our execution time expectations, we assume that they are of case (ii), caused by the under-provisioned cloud behind the virtual infrastructure executing the observed jobs.

Below, we present a few workflow execution time error functions that match the above criteria. Later, if we refer to a particular function from below, we will use one of the subscripted versions of $E$, otherwise, when the particular function is not relevant, we just use $E$ without subscript. Although the algorithm and techniques discussed later are independent from the applied $E$ function, these functions are not interchangeable, their error values are not comparable at all.

\begin{description}
	\item[\emph{Average distance.}] This error function calculates the average time discrepancy of the first $k$ jobs.
	
{\small
\begin{equation}\label{EQ-SQD}
E_{SQD}(W,t,k):=\sqrt{\frac{\sum_{1\leq i\leq k} (r_{ex}(j_i)-r_{ob}(j_i,t))^2}{k}}
\end{equation}
}

\item[\emph{Mean absolute percentage error.}] Here the relative error of the observed runtime is calculated for each job, then it is averaged for all $k$ jobs:

{\small
\begin{equation}\label{EQ-MAPE}
E_{MAPE}(W,t,k):=\frac{100}{k}\sum_{1\leq i \leq k}\frac{|r_{ex}(j_i)-r_{ob}(j_i,t)|}{r_{ex}(j_i)}
\end{equation}
}

\item[\emph{Time adjusted distance.}] The function adjusts the execution time discrepancies calculated in $E_{SQD}$ so that the jobs started closer (in time) to $j_k$ will have more weight in the final error value. 

{\small\begin{equation} \label{EQ-TADJ}
E_{TAdj-SQD}(W,t,k):=\sqrt{\frac{\sum_{1\leq i\leq k} \frac{i}{k}(r_{ex}(j_i)-r_{ob}(j_i,t))^2}{\sum_{1\leq i\leq k}\frac{i}{k}}}
\end{equation}}
\end{description}

\section{Workflow enactment and simultaneous prediction}
\label{sec:core}

\begin{figure*}
	\centering
	\includegraphics[width=0.8\textwidth]{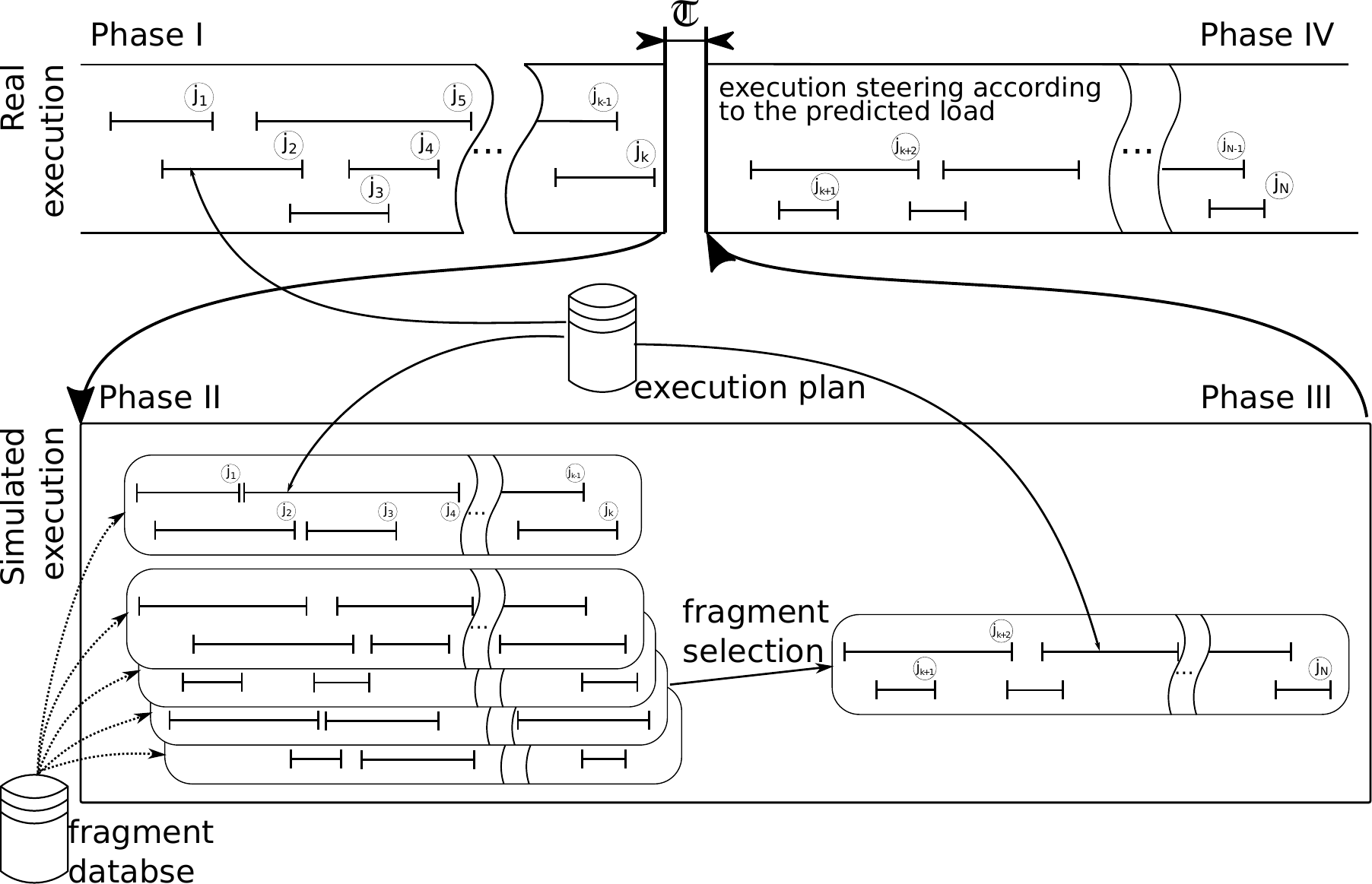}
	\caption{Phases of our workload prediction in relation to the workflow being ran\label{FIG-SCHEME}}
\end{figure*}

When job $j_k$ is completed during the execution (phase I in Figure~\ref{FIG-SCHEME}), a deviation analysis is performed using one of the error functions of Eq.~\ref{EQ-SQD}-\ref{EQ-TADJ}  to compare the actual job execution times to the ones in the enactment plan. Significant deviations -- $E(W,t,k)>E_\epsilon$, where $E_\epsilon$ is predefined by the workflow developer -- initiate the background workload prediction phase that corresponds to the second, Analysis phase of autonomous control loops. This phase is omitted, if the workflow enactor estimates the remaining workflow execution time is smaller than required for background workload prediction. The maximum time spent on workload prediction is limited by a predefined $\mathfrak{T}$, represented as a gap in the execution in Figure \ref{FIG-SCHEME}. Thus, workload prediction is not performed if $\sum_{i=k+1}^N r_{ex}(j_i)<\mathfrak{T}$.

\subsection{Background workload prediction} \label{SEC-BWP}

In essence, we simulate the workflow execution on a given cloud infrastructure while adding {\em known} workloads as background load (phase II in Figure \ref{FIG-SCHEME}). The workflow is simulated according to the enactment plan specified runtime properties, like job start time, completion time, and times for creating virtual machines. We expect the simulated workflow to match its real-world counterpart (in terms of runtime properties), when the added background load closely estimates the real-world load. We use Eq.~\ref{EQ-SQD}-\ref{EQ-TADJ} to find the known workload closest to the observed one (note: only one function should be used during the whole prediction procedure). Next, we present the details of Algorithm~\ref{ALG-GFBP} that implements this background workload matching mechanism.

\subsubsection{Base definitions.}
Before diving into the details, we provide a few important definitions: \emph{trace fragment} -- used to provide a particular background workload~--, \emph{past error} -- determines the previously collected execution time error values regarding the completed jobs $j_1, ...j_k$ -- finally, \emph{future error} -- defines the previously collected execution error evaluations of the jobs $j_{k+1},... j_N$ that have not yet run in the current workflow instance (i.e., what was the level of error in the ``future'' that after a particular past error value was observed).

A \emph{trace fragment} is a list of activities characterised by such runtime properties (e.g., start time, duration, performance, etc.) that are usable in simulators, their collection is denoted as fragment database in Figure \ref{FIG-SCHEME}. Each fragment represents realistic workloads i.e., real-world system behaviour. Fragments are expected to last for the duration of the complete simulation of the workflow with all its jobs. The fragment duration is independent from the actual real life situation modelled -- which stems from the actual $[j_k,t]$ job which triggered the prediction. In a worst case, fragments should last for the completely serial execution of the workflow: $\sum_{i=1\dots k}r_{ob}(j_i)+\sum_{i=k+1\dots N}r_{ex}(j_i)$. Thus fragment durations vary from workflow-to-workflow. Apart from their duration, fragments are also characterised and identified by their starting timestamp, i.e. the time instance their first activity was logged, denoted as $t \in T$ (where $T\subset\mathcal{T}$); later we will refer to particular fragments by their identifying starting timestamp (despite these fragments often-times contain thousands of activities). As a result, when the algorithm receives an identifying timestamp, it queries the trace database for all the activities that follow the first activity for the whole duration of the fragment. Note, that our algorithm uses the relative position of these timestamps. Therefore, when storing historic traces as fragments, they are stored so that their timestamps are consistent and continuous, this requires some displacement of their starting positions. This guarantees that we can vary the fragment boundaries (according to the workflow level fragment duration requirements) at will.

Arbitrary selection of fragment boundaries would result in millions of trace fragments. If we would simulate with every possible fragment, the analysis of a single situation would take days. However, as with any prediction, the longer time it takes the less valuable its results become (as the predicted future could turn past by then). Predictions typically are only allowed to run for a few minutes as a maximum, thus the entire simulation phase must not hinder the real execution for more than $\mathfrak{T}$. In the following, we survey the steps that are necessary to meet this requirement i.e., reducing the analysis time from days to $\mathfrak{T}$. The fragment database needs to be {\em pre-filter}ed so only a few fragments ($T_{filt}\subset T$) are used in the analysis later on. 

Although this is out of scope of the current article, pre-filtering can use approaches like pattern matching, runtime behaviour distance minimisation (e.g., by storing past workflow behaviour -- for particular fragments -- and by comparing to the current run to find a likely start timestamp), or even random selection. Filtering must take limited and almost negligible time. In our experiments, we assumed it below $\mathfrak{T}/1000$ (allowing most of the time to be dedicated to the simulation based analysis of the situation). As a result of filtering, the filtered fragment count must be reduced so that the time needed for subsequent simulations does not exceed the maximum time for predictions: $|T_{filt}|<\mathfrak{T}/\mathfrak{t_{sim}}(W)$, where $\mathfrak{t_{sim}}$ is the mean execution time of $W$ in the simulated cloud. Our only expectation that the pre-filtering happens only in memory and thus the fragment database is left intact. As a result, when we run the algorithm, it only sees a portion of the fragment database. On the other hand, future runs of the algorithm might get a different portion from the database depending on the future runtime situation.

Finally, we dive into the \emph{error definitions}. Alongside fragments, several error values are also stored in the fragment database, but unlike fragments, which are independent from workflows, these error values are stored in relation to the particular workflow and its already completed instances. Later, just like projected execution times, these stored error values are also going to be used to steer the algorithm. First, in terms of \emph{past errors} we store the values received from our previously defined partial execution time error functions, E.q.~\ref{EQ-SQD}-\ref{EQ-TADJ}. Past errors are stored for every possible $k$ value for the particular workflow instance. We also calculate future errors similarly to past errors. Our calculation uses the part of the workflow containing the jobs after $j_k$:~$W^F(W,k):=\{\forall j_i\in W:i>k\land i\leq N\}$, where $W^F\in \mathcal{W}$. Thus, the future error function determines how a particular, previously executed workflow instance continued after a specific past error value:
\begin{equation}\label{EQ-FERR}
F(W,t,k):=E(W^F(W,k),t,N-k).
\end{equation}
This function allows the evaluation and storage of the final workflow execution time error for those parts of the past workflow instances, which have not been executed in the current workflow execution.

\subsubsection{Overview of the algorithm.}

\begin{figure}[tb]
\centering
\includegraphics[width=\columnwidth]{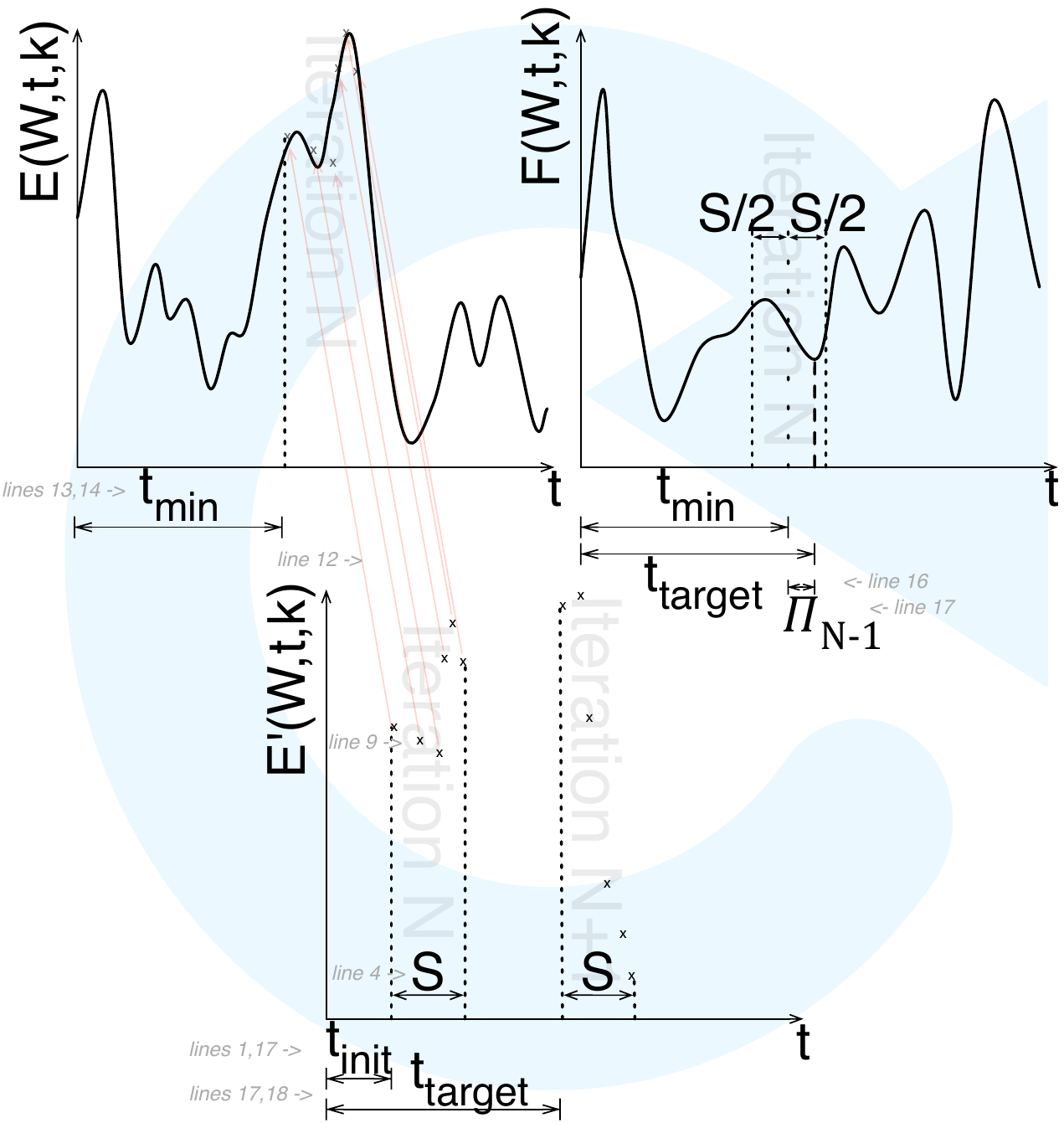}
\caption{Visual representation of the algorithm\label{fig:visualalgo}}
\end{figure}

\begin{algorithm}[!tb]
\small
\begin{algorithmic}[1]
\REQUIRE $T_{filt}\subset T$ -- the filtered trace fragment set
\REQUIRE $S \in \mathbb{R}^+$ -- the primary search window size
\REQUIRE $\Pi \in\mathbb{R}^+$ -- the precision of the trace match
\REQUIRE $I \in \mathbb{N}$ -- the maximum iteration count 
\REQUIRE $P \in \mathbb{R}^+$ -- max evaluations for searching in function $\phi(x)$
\REQUIRE $[W, t_{curr}]$ -- the current workflow instance
\REQUIRE $R_{ex}:=\{r_{ex}(j_i):\{j_i\in W\}\}$ -- the model execution times 
\REQUIRE $R_{ob}:=\{r_{ob}(t_{curr},j_i):\{j_i\in W\land i\leq k\}\}$  -- the observed execution times 
\ENSURE {$t_{target}$ is around the approximated workload}

\STATE $t_{init} \gets t\in T_{filt}$ \label{ALG-LININIT} 
\STATE $T_{list} \gets \emptyset$ 
\REPEAT 
	\STATE $\mathfrak{I}\gets (t_{init}-S/2,t_{init}+S/2)$
	\STATE $\mathfrak{R} \in 2^\mathfrak{I}\backslash \{\emptyset\}$ -- arbitrary choice \label{ALG-LINSEASP}
	\FORALL {$t\in \mathfrak{R}$} 
		\FORALL {$j_i\in W:i<k$} 
			\STATE $r_{ex}'(j_i)\gets r_{ob}(t_{curr},j_i)$  \label{ALG-LINEXSUBST}
			\STATE $r_{ob}'(j_i,t)\gets sim(W, R_{ex}, i, t)$ \label{ALG-LINSIM} 
		\ENDFOR
	\ENDFOR
	\STATE $T_{red} \in 2^{T \backslash T_{list}} : |T_{red}|=P$ \label{ALG-TRED}
	\STATE $T_{min} \gets \{t\in T |\phi(t)=\underset{x\in \{T_{red}\}}{\min}\phi(x)\}$ 
	\STATE $t_{min} \gets \min T_{min}$\label{ALG-LINTMIN}
	\STATE $T_{list} \gets T_{list}\cup\{t_{min}\}$ \label{ALG-LINPOPTM}
	\STATE $t_{target}(|T_{list}|) \gets \{t_l\in T:F(W,t_l,k)-E(W,t_l,k)=\underset{{t_{min}-S/2<t<t_{min}+S/2}}{\min} (F(W,t,k)-E(W,t,k))\}$ \label{ALG-LINFUTMIN}
	\STATE $t_{init}\gets t_x\in T_{filt}:|t_x-t_{target}(|T_{list}|)|=min_{t\in T_{filt}}|t-t_{target}(|T_{list}|)|$ \label{ALG-LINALIGN}
 \UNTIL {$(|t_{target}(|T_{list}|)-t_{target}(|T_{list}|-1)|>\Pi)\land(|T_{list}|<I)$} \label {ALG-EXITCOND}

\RETURN $t_{tareget}(|T_{list}|)$
\end{algorithmic}
\caption{Fitting based prediction\label{ALG-GFBP}}
\end{algorithm}

Algorithm~\ref{ALG-GFBP} (also depicted functionally in Figure~\ref{fig:visualalgo} and structurally as phases II-III in Figure \ref{FIG-SCHEME}) aims at finding a timestamp so that the future estimated error is minimal, while past error prediction for this timestamp is the closest to the actual past error (i.e., the estimated and actual ``past errors are aligned''). The algorithm is based on the assumption that if past workloads are similar (similarity measured by their error functions) then future workloads would be similar, too.

In detail, line \ref{ALG-LININIT} picks randomly one of the fragments identified by the timestamp in the filtered set $T_{filt}$  and stores in $t_{init}$. This will be the assumed initial location of the fragment that best approximates the background load. Later, in line \ref{ALG-LINALIGN}, this $t_{init}$ will be kept updated so it gives a fragment that  better approximates the background load. The \emph{primary search window} -- $\mathfrak{R}$ of line \ref{ALG-LINSEASP} also shown between the dashed lines of the lower chart in Figure~\ref{fig:visualalgo} -- represents a set of timestamps within a $S/2$ radius from the assumed start of the fragment specified by $t_{init}$. The algorithm uses set $T_{list}$ to store timestamps for the approximate trace fragments  as well as to count the iterations (used after line \ref{ALG-LINPOPTM}).

A simulator is used to calculate observed  execution times $r_{ob}'$ for the jobs in the simulated infrastructure (see line \ref{ALG-LINSIM}). This is expressed with $sim(W,R_{ex},i,t)$ thus, each simulation receives the workflow to be simulated, the set of execution time expectations ($R_{ex}$) that specify the original enactment plan, the identifier of the job ($1 \leq i \leq N$) we are interested in and the timestamp of the trace fragment ($t\in T$) to be used in the simulation as background load to the workflow, respectively. With these parameters the simulator is expected to run all the activities in the trace fragment identified by $t$ in parallel to a simulated workflow instance. Note: the simulation is done only once for the complete workflow for a given infrastructure and a given fragment, later this function simply looks up the past simulated $r_{ob}'$ values.

Next, we use one of the error functions of the workflow execution time as defined in Eq.~\ref{EQ-SQD}-\ref{EQ-TADJ}. As we have simulated results, we substitute the observed/expected execution time values in the calculation with their simulated counterparts. In the simulation, the expected execution times $r_{ex}'$ are set as the real observed execution times $r_{ob}$ (see line~\ref{ALG-LINEXSUBST}). To denote this change in the inputs to the error function, we use the notation of $E'(W,t,k)$ --  error of simulated execution time. This function shows how the simulated workload differs from the  observed one. The evaluation of the $E'$ function is depicted with $\times$ marks at the bottom chart of Figure~\ref{fig:visualalgo}.

Afterwards, lines~\ref{ALG-TRED}-\ref{ALG-LINTMIN} search through the past error values for each timestamp (using the same error function as we used for the evaluation of $E'$). With the help of function $\phi: \mathcal{T} \to \mathbb{R} $:
\begin{equation}
\phi(x):=\sum_{t\in \mathfrak{R}}\big|E'(W,t,k)-E(W,x+t-t_{init}+S/2,k)\big|
\end{equation} 
This function offers the difference between the simulated and real past error functions (Figure~\ref{fig:visualalgo} represents this with red projection lines between the chart of $E$ and the $\times$ marks of $E'$). The algorithm uses the $\phi(x)$ function to find the best alignment between the simulated and real past error functions: we set $t_{min} $ as the time instance in $\mathfrak{R}$ with the smallest difference between the two error functions. The alignment is searched over an arbitrary subset of the timestamps: $T_{red}$ -- the \emph{secondary search window}. The algorithm selects a $T_{red}$ with a cardinality of $P$ in order to limit the time to search for $t_{min}$. The arbitrary selection of $T_{red}$ is used to properly represent the complete timestamp set of $T$. 

After finding $t_{min}$, we have a timestamp from the fragment database, for which the behaviour of the future error function is in question, this corresponds to the fragment selection in Figure~\ref{FIG-SCHEME}. Line~\ref{ALG-LINFUTMIN} finds the timestamp that has the closest past and future error values in the range around $t_{min}$ within radius $S/2$ -- see also in the top right chart of Figure~\ref{fig:visualalgo}. Note, this operation utilises our assumption that past and future errors are aligned (ie., a trace fragment with small past error value is more likely to result in similarly small future error value). The timestamp with the future error value closest to the past error is used as $t_{target}$ for the current iteration (i.e., our current estimate for the start timestamp of the approximate background load).

Finally, the iteration is repeated until te successive change in $t_{target}$ is smaller than the precision $\Pi$ or the iteration count reaches its maximum -- $I$, represented as phase III in Figure \ref{FIG-SCHEME}. Note, $I$ is set so the maximum time spent on workload prediction ($\mathfrak{T}$) is not violated. The algorithm then returns with the last iteration's $t_{target}$ value to represent the starting timestamp of the predicted trace fragment that most resembles the background load currently experienced on the cloud behind the workflow. This returned value (and the rest of the trace fragment following $t_{target}$) then could be reused by when utilizing the real life version of the simulated cloud. For example, the workflow enactor could use the knowledge of the future expected workload for the planning and execution phases of its autonomous control loop (phase IV in Figure \ref{FIG-SCHEME}). Note, the precise details on the use of the predicted workload is out of the scope of this article.

\section{Evaluation with a Biochemical Workflow}
\label{sec:eval}
\begin{figure*}[tb]
	\centering
	\includegraphics[width=\textwidth]{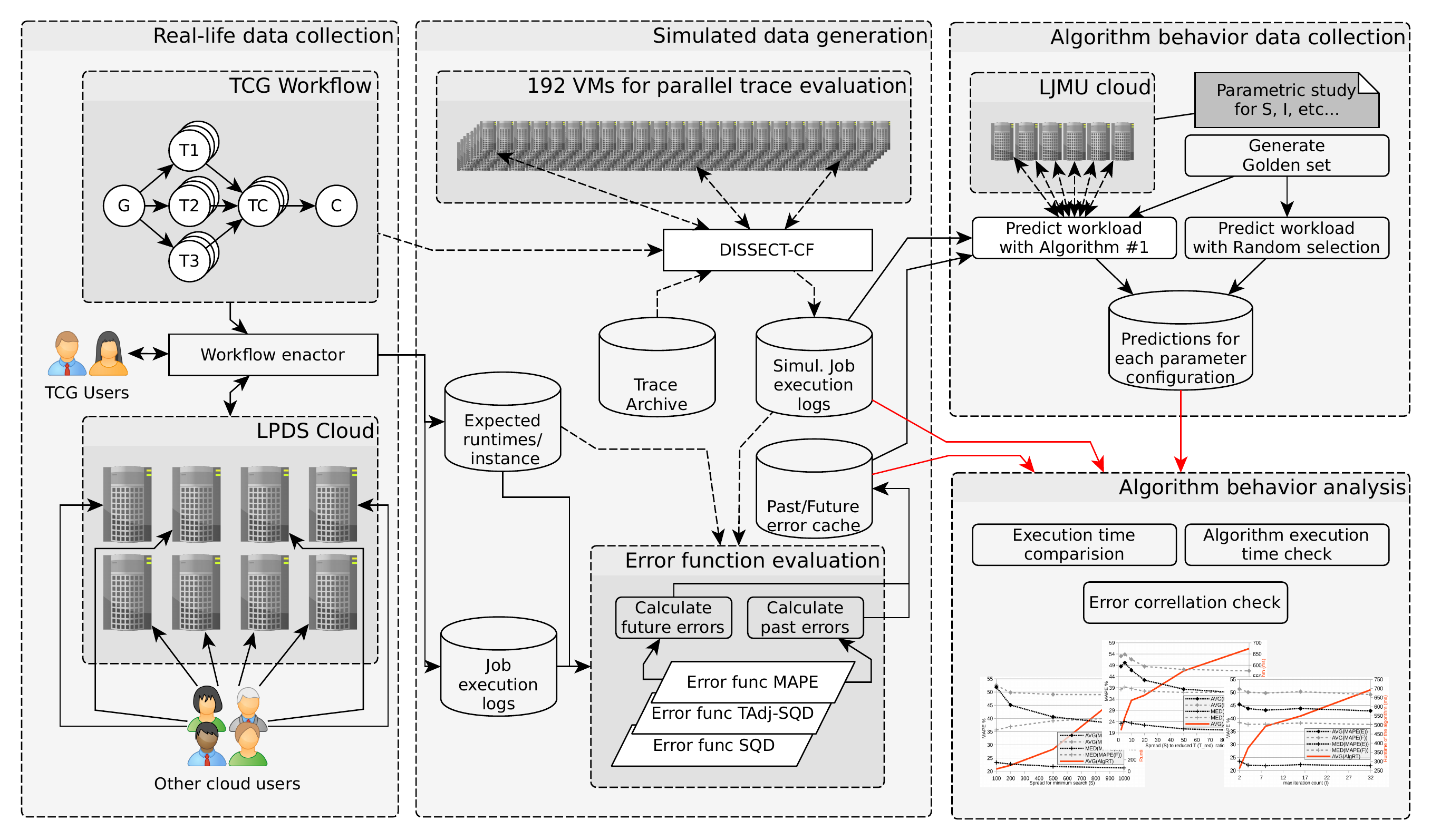}
	\caption{Our detailed evaluation approach\label{fig:compvalid}}
\end{figure*}

We demonstrate our approach via a biochemical workflow that generates conformers by unconstrained molecular dynamics at high temperature to overcome conformational bias then finishes each conformer by simulated annealing and/or energy minimisation to obtain reliable structures. It uses the TINKER library \cite{tinker_ccpe} for molecular modelling for QSAR studies for drug development. 

Our evaluation approach is summarized in Figure~\ref{fig:compvalid}. It is composed of three main phases: $(i)$ data collection from a real life environment, $(ii)$ modelling the TCG workflow and simulating its behaviour under various background loads, and $(iii)$ evaluating the algorithm based on the collected real life and simulated data.

\subsection{The Tinker workflow on the LPDS cloud} \label{sec:TCGONLPDS}

\begin{figure*}[tb]
	\centering
	\begin{framed}
		\footnotesize
		\begin{verbatim}
 1|  PSSTART
 2|  VMDEF VA=tinker,25,0,306176000 RC=1,5.0E-4,1073741824 VAST=iscsi-izabel DATA=iscsi-izabel
 3|  #This is the seq for job G
 4|  VMSEQ N50500 N190000 N334000 C22.333 N4578744 C0.5 N6150209 N1488 C43200
 5| 
 6|  PSSTART
 7|  # VM instance description for the VMs to be used
 8|  VMDEF VA=tinker,25,0,306176000 RC=1,5.0E-4,1073741824 VAST=iscsi-izabel DATA=iscsi-izabel
 9|  # 20x the main PS part of the workflow
10|  # The last four C-s are jobs T1, T2, T3 and TC from the workflow
11|  VMSEQ N50500 N190000 N334000 C22.333 N4578744 C0.5 N6150209 N1488 C2145 C3573 C1886 C2
12|  VMSEQ N50500 N190000 N334000 C22.333 N4578744 C0.5 N6150209 N1488 C2145 C3573 C1886 C2
13|  #                     [...]
14|  #                     [... this VMSEQ is repeated for 17x (resulting in 20 VMs in parallel)]
15|  #                     [...]
16|  VMSEQ N50500 N190000 N334000 C22.333 N4578744 C0.5 N6150209 N1488 C2145 C3573 C1886 C2
17|
18|  #                     [... the above PSSTART section repeated for 14 more times]
19|
20|  PSSTART
21|  VMDEF VA=tinker,25,0,306176000 RC=1,5.0E-4,1073741824 VAST=iscsi-izabel DATA=iscsi-izabel
22|  #This is the seq for job C
23|  VMSEQ N50500 N190000 N334000 C22.333 N4578744 C0.5 N6150209 N1488 C10.6
\end{verbatim}
	\end{framed}
	\caption{The description of the TCG workflow's execution for the simulation\label{FIG-WF-DESC}}
\end{figure*}

The TINKER Conformer Generator (TCG) workflow \cite{tinker_ccpe} consists of 6 steps (see the top left corner of Figure~\ref{fig:compvalid}):  $(i)$ \textbf{G}: generating 50000 input molecule conformers (taking around 12 hours, compressed into 20 zip files by grouping 2500 conformers); $(ii)$ \textbf{T1}: minimising the initial conformational states generated at high temperature; $(iii)$ \textbf{T2}: performing a short low temperature dynamics with the high temperature conformations to simulate a low temperature thermodynamic ensemble, and minimising the above low temperature states; $(iv)$ \textbf{T3}: cooling the high temperature states by simulated annealing, and minimising the annealed states; $(v)$ \textbf{TC}: collecting parameter study results; $(vi)$ \textbf{C}: re-compressing results to a single file. The sequential execution of the workflow on a single core 2 GHz CPU takes around 160 hours. 

Note that the parallel section of the workflow could be partitioned arbitrarily: by changing how \textbf{G} splits the input molecule conformers. For example, any options could be chosen from the extreme case of each molecule becoming an input for a separate \textbf{T1/2/3/C} run, to the other extreme of having a single zip file produced and processed in a single \textbf{T1/2/3/C}. Also, if a larger infrastructure is available, more input molecule conformers could be considered in a single run (this would result in a longer execution time for \textbf{G}).

In the period of over half a year, the workflow was ran several times on the cloud of the Laboratory of Parallel and Distributed Systems -- LPDS cloud --, which ran OpenNebula 4.10 and consisted of 216 cores, 604 GBs of memory and 70 TBs of storage at the time of the experiments. We used a workflow enactor without autonomous control mechanisms. 
We have collected the job execution times for all jobs in the workflow, as well as the  time instance when the workflow was started. We have calculated the expected job execution times -- $r_{ex}(j_i)$ -- as an average of the execution times observed. This average was calculated from over 500 runs for each step of the TCG workflow. To enable a more detailed analysis of the workflow executions, we have generated larger input sets allowing us to repeatedly execute the parallel section with 20 virtual machines (in our implemented workflow, the 20 machine parallel section was executed 15 times before concluding with the final re-compression phase -- \textbf{C}).
This allowed us to populate our initial past and future error values in the cache (ie., we have calculated how particular workflow instances behaved when expected job execution times are set to be the average of all). Not only the error cache was populated though, the individual $r_{ob}(j_i,t)$ values were also was stored in our database (in total, the collected data was about 320MBs). These data stores are shown in Figure~\ref{fig:compvalid} as a \emph{per instance expected runtimes} database,  \emph{job execution logs}) and \emph{Past/Future error cache}. The stored values acted as the foundation for the simulation in the next phase of our evaluation.

\subsection{Modelling and simulating the workflow} 

This sub-section provides an overview on how the TCG workflow was executed in a simulator. Our choice for the simulator was the open source DISSECT-CF\footnote{\url{https://github.com/kecskemeti/dissect-cf}}. We have chosen it because it is well suited for simulating resource bottlenecks in clouds, it has shown promising performance gains over more popular simulators (e.g., CloudSim, SimGrid) and its design and development was prior work of the authors~\cite{SIMPAT15}. The sub-section also details the captured properties of the TCG workflow, which we collected in previous phase of the evaluation. Then, as a final preparatory step for our evaluation, we present the technique we used to add arbitrary background load to the simulated cloud that is used by the enactor to simulate the workflow's run.  

\subsubsection{The model of the workflow's execution}
The execution of the TCG workflow was simulated according to the description presented in Figure~\ref{FIG-WF-DESC}. The description is split into three main sections, each starting with a \verb+PSSTART+ tag (see lines 1, 6 and 20, which correspond to the three main sections G--[T1/T2/T3--TC]--C of the TCG workflow shown in the top left corner of Figure~\ref{fig:compvalid}). This tag is used as a delimiter of parallel sections of the workflow, thus everything that reside in between two \verb+PSSTART+ lines should be simulated as if they were executed in parallel. Before the actual execution though, every \verb+PSSTART+ delimited section contains the definition of the kind of VM that should be utilized during the entire parallel section. The properties of these VMs are defined by the \verb+VMDEF+ entry (e.g., see lines 2 or 8) following \verb+PSSTART+ lines. Note, the definition of a VM is dependent on the simulator used, so below we list the defining details specific to DISSECT-CF:
\begin{description}
	\item[\emph{The virtual machine image}] used as the VM's disk. This is denoted with property name \verb+VA+. In this property, we specify that the image is to be called ``\emph{tinker}''. Next, we ask its boot process to last for 25 seconds. Afterwards, we specify the VM image to be copied to its hosting PM before starting the VM  -- 0 (i.e., the VM should not run on a remote filesystem). Finally, we set the image's size as 306 MBs. 
	\item[\emph{The required resources}] to be allocated for the VM on its hosting PM. These resources are depicted behind the property name of \verb+RC+ in the figure. Here we provided details for the number of cores (1), their performance (\verb+5.0E-4+ -- this is a relative performance metric compared to one of the CPUs in LPDS cloud) and the amount of memory (1 GB) to be associated with the soon to be VMs.
	\item[\emph{Image origin}] where the VM's disk image is downloaded from before the virtual machine is instantiated. We used the property name of \verb+VAST+ to tell the simulator the host name of the image repository that originally stores the VM's image.
	\item[\emph{Data store}] is the source/sink of all the data the VM produces during its runtime. This is defined with the property called \verb+DATA+.  This field helps the simulation to determine the target/source of the network activities later depicted in the \verb+VMSEQ+ entries.
\end{description}
The real-life workflow was executed in the LPDS cloud (see the leftmost section of Figure~\ref{fig:compvalid}). Thus, we needed to model this cloud to match the simulated behaviour of the workflow to its real life counterpart ran in phase one. Therefore, the storage name $iscsi-izabel$ in the workflow description (e.g., in line 2 of Figure~\ref{FIG-WF-DESC}) refers to the particular storage used on LPDS cloud, just like the VMI image name $tinker$ does.

Now, we are ready to describe the runtime behaviour of the workflow observed in phase one in a format easier to process by the simulation. This behaviour is denoted with the \verb+VMSEQ+ entries (e.g., see lines 4 or 10) that reside in each \verb+PSSTART+ delimited parallel section. \verb+VMSEQ+ entries are used to tell the simulator a new VM needs to be instantiated in the parallel section. Each VM requested by the \verb+VMSEQ+ entries will use the definition provided in the beginning of the parallel section. All VMs listed in the section are requested from the simulated LPDS cloud right before the workflow's processing reaches the next \verb+PSSTART+ entry in the description. This guarantees they are requested and executed in parallel (note, despite requesting the VMs simultaneously from the cloud, their level of concurrency observed during the parallel section will depend on the actual load of the simulated LPDS cloud). The processing of the next parallel section, only starts after the termination of all previously created VMs. 

In the \verb+VMSEQ+ entries, a VM's activities before termination. There are two kinds of activities listed: network and compute. Network activities start with \verb+N+ and then followed by the number of bytes to be transferred between the \verb+DATA+ store and the VM (this is the store defined by the \verb+VMDEF+ entry at the beginning of the parallel section). Compute activities, on the other hand, start with the letter \verb+C+ and then they list the number of seconds till the CPUs of the VM are expected to be fully utilised by the activity. VM level activities are executed in the simulated VM in a sequence (i.e., one must complete before the next could start). 

For example, line 12 of Figure~\ref{FIG-WF-DESC} defines how job executions are performed in a VM. First, we prepare the VM to run the tinker binaries by installing three software packages. This results in three transfers (49 KB, 186 KB and 326 KB files) and a task execution for 22 seconds. Next, we fetch the $tinker$ package (4.4 MBs) and decompress it (in a half a second compute task). Then, we transfer the input files with the 2500 conformers and the required runtime parameters to use them (5.9 MBs and 1.5 KBs). Afterwards, we execute the T1, T2, T3 and TC jobs sequentially taking 35, 60, 32 minutes and 2 seconds, respectively. These values were gathered as the average execution times for the jobs while the real life workflows were running in LPDS cloud. Finally, this 2 second activity concludes the VMs operations, therefore it is terminated.

The \verb+PSSTART+ entry in line 6 and the virtual machine executions defined until line 16 represent a single execution of the parallel section of the TCG workflow. 
Because of repetitions, we have omitted the several \verb+VMSEQ+ entries from the parallel section, as well as several  \verb+PSSTART+ entries representing further parallel sections of conformer analysis. On the other hand, the description offered for the simulator did contain all the 14 additional \verb+PSSTART+ entries which were omitted here for readability purposes. 

To conclude, the description in Figure~\ref{FIG-WF-DESC} provides details for over 1800 network and computing activities to be done for a single execution of the TCG workflow. If we consider only those activities that are shown in the TCG workflow, we still have over 1200 computing activities remaining. These activities result in the creation and then destruction of 302 virtual machines in the simulated cloud. When calculating the error functions, we would need expected execution details for all these activities or VMs. The rest of the article will assume that the workflow enactor provides details about the computing activities directly relevant for the TCG workflow only (i.e., the jobs of \textbf{G}/\textbf{T}\emph{x}/\textbf{C}). Thus our $N$ value was 1202. The partial workflow execution error functions could be evaluated for every job done in the simulated TCG. This, however, is barely offering any more insight than having an error evaluation at the end of each parallel section (ie., when all VMs in the particular parallel section are complete). As a result, in the rest of the article, when we report $k$ values, they are going to represent the amount of parallel sections complete and not how many actual activities were done so far. To transform between activity count and the reported $k$ values one can apply the following formula:
\begin{equation}
k_{real}:=\left\lbrace \begin{tabular}{ll}k=0&0\\
k\textless 15&1+80k\\
k=15&1+80k+1
\end{tabular}\right.
\end{equation} 
,where 15 is the number of parallel sections, and 80 is the number of TCG activities per parallel section. Finally, the $k_{real}$ is the value used in the actual execution time error formulas from Eq.~\ref{EQ-SQD}-\ref{EQ-TADJ}.

Although, the above description was presented with our TCG workflow, the tags and their attributes of the description were defined with more generic situations in mind. In general, our description could be applied to workflows and applications that have synchronisation barriers at the end of their parallel sections. 

\subsubsection{Simulating the background load}
\begin{figure*}[tb]
	\centering
	\includegraphics[width=0.85\textwidth]{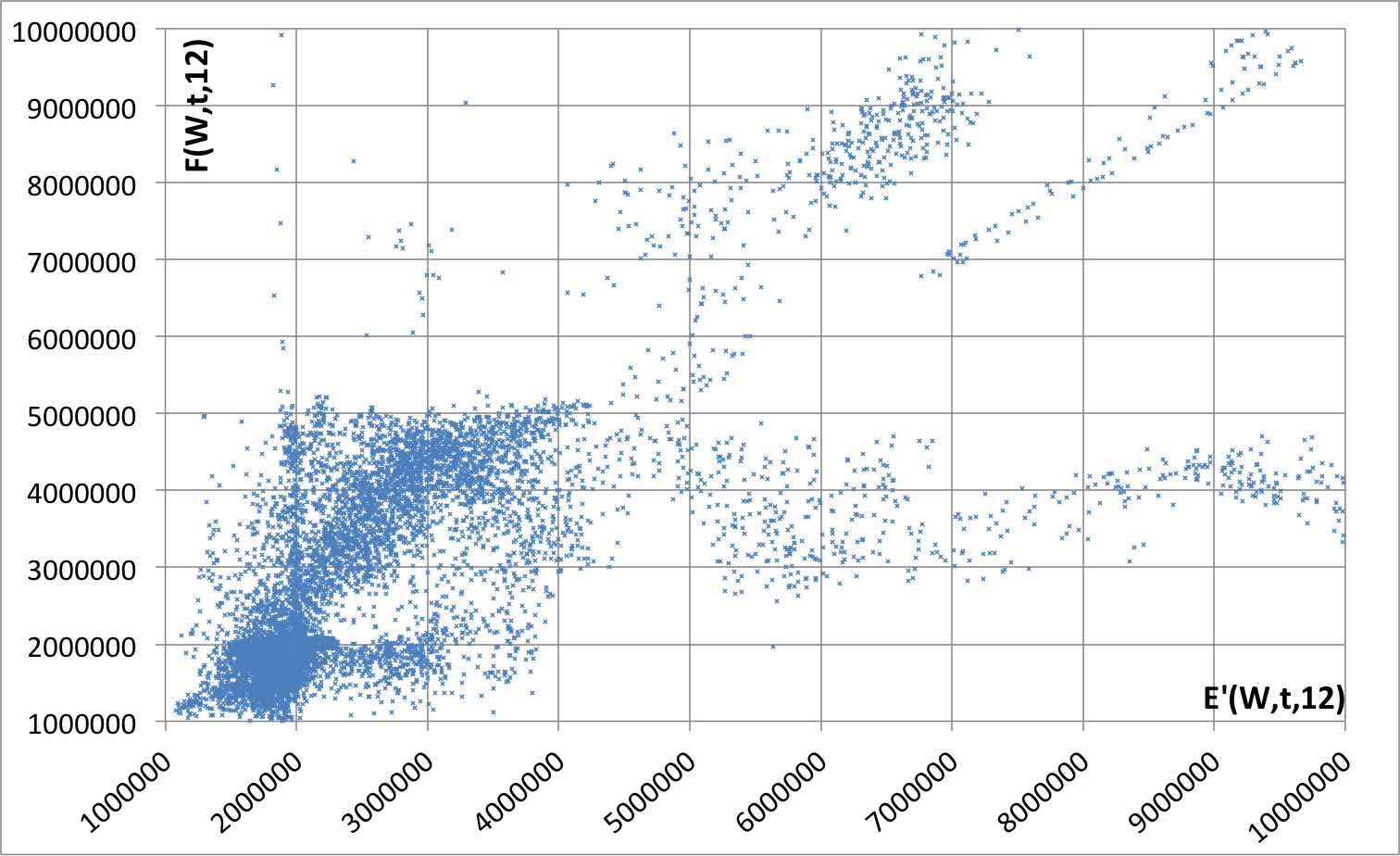}
	\caption{Nordugrid trace as the background load behind TCG, exemplified with the use of  error function $E_{SQD}$}\label{fig:exnordu}
\end{figure*}

In order to simulate how the workflow instances of TCG would behave under various workload conditions, we needed a comprehensive workload database. We have evaluated previously published datasets: we looked for workload traces that were collected from scientific computing environments (as that is more likely to resemble the workloads behind TCG). We only considered those traces that have been collected over the timespan of more than 6 months (ie., the length of our experiment with TCG on the LPDS cloud -- as detailed in Section~\ref{sec:TCGONLPDS}). We further filtered the candidate traces to only contain those which would not cause significant (i.e., months) overload or idle periods in the simulated LPDS cloud. This essentially left 4 traces (SharcNet, Grid5000, NorduGrid and AuverGrid) from the Grid Workloads Archive (GWA~\cite{Iosup08thegrid}), we will refer to the summary of these traces as $T^{GWA}\subseteq T$. Note, it is irrelevant that the traces were recorded on grids -- for our purposes the user behaviour, i.e., the variety of tasks, their arrival rate and duration are important. Fortunately these characteristics are all independent from the actual type of infrastructure.

Trace fragments were created using GWA as follows: we started a fragment from every job entry in the trace. Then we identified the end of the fragment as follows. Each fragment was expected to last at least for the duration of the actual workflow instance. Unfortunately, the simulated LPDS cloud could cause distortions to the job execution durations (e.g., because of the different computing nodes or because of the temporary under/over-provisioning situations compared to the trace's original collection infrastructure), making it hard to determine the exact length of a fragment without simulation. Thus, we have created fragments that included all jobs within 3 times the expected runtime of the workflow. As a result each of our fragments was within the following time range: $[t,t+3\sum_{0<i<N}r_{ex}(j_i)]$. All these fragments were loaded in the \emph{trace archive} of Figure~\ref{fig:compvalid} (see its middle section titled ``simulated data generation''). All together, our database have contained more than 2 million trace fragments. 

With the trace fragments in place, we had every input data ready to evaluate the $r'_{ob}$ values under various workload situations (ie., represented by the fragments). Thus, we have set out to create a large scale parametric study. For this study, we have accumulated as many virtual machines from various cloud infrastructures as many we could afford. In total, we have had 3 cloud infrastructures (SZTAKI cloud\footnote{\url{http://cloud.sztaki.hu/en}}, LPDS cloud, Amazon EC2\footnote{\url{https://aws.amazon.com/ec2}}) involved which hosted 192 single core virtual machines with 4 GBs of memory and 5 GBs of hard disk storage (and the closest equivalent on EC2). We offered the trace archive to all of them as a network share. Each VM hosted DISSECT-CF v0.9.6 and was acting as a slave node for our parametric study. The master node (not shown in the figure), then instructed the slaves to process one trace at a time as follows\footnote{The source code of these steps are published as part of the following project: \url{https://github.com/kecskemeti/dissect-cf-examples/}}:
\begin{enumerate}
	\item Load a trace fragment as per the request of master.
	\item Load the description of LPDS cloud\footnote{\url{http://goo.gl/q4xZpe}}.
	\item Load the description of TCG workflow execution (ie., the one shown in Figure~\ref{FIG-WF-DESC}).
	\item Start to submit the jobs from the loaded fragment to the simulated LPDS cloud (for each submitted job, our simulation asks a VM from the cloud which will last until the job completes).
	\item Wait until the 50th job -- this step ensures the simulated infrastructure is not under a transient load.
	\item Start to submit the jobs and virtual machines of the workflow execution specified in the previously loaded description (the VMs here were also ran in the simulated LPDS cloud).
	\item For each task, record its observed execution time -- $r'_{ob}(j_i,t)$. Note, here $t$ refers to the start time of the simulated workflow.
	\item After the completion of the last job and VM pair in the workflow,  terminate the simulation.
	\item Send the collected job execution times to master.
\end{enumerate}
The simulation of all trace fragments took less than 2 days. The mean simulation execution time for a fragment running on our cloud's model is $\mathfrak{t_{sim}}(TCG)=756ms$. We have stored the details about each simulation in relation to the particular trace fragment in our simulated job execution log database (see Figure~\ref{fig:compvalid}). 

To conclude our simulated data generation phase, we populated our \emph{past and future error cache} of Figure~\ref{fig:compvalid}. Later our algorithm used this to represent past workflow behaviour. We calculated the past and future error values with the help of all $r'_{ob}$ we collected during the simulation phase. The error values were cached from all 3 error functions we defined in Eq.~\ref{EQ-SQD}-\ref{EQ-TADJ} as well as from their future error counterparts from Eq.~\ref{EQ-FERR}. This cached database allowed us to evaluate the algorithm's assumptions and behaviour in the simulated environment as we discuss it in the next subsection.

\subsection{Evaluation}


In this subsection, we evaluate our algorithm using the collected data about the simulated and real life TCG workflow instance behaviour. We focused on three areas: $(i)$ analyse our assumption on the relation of past and future errors, $(ii)$ provide a performance evaluation of the algorithm, and $(iii)$ analyse how the various input variables to the algorithm influence its accuracy. These are shown as the last two phases (\emph{algorithm behaviour data collection and analysis}) in Figure~\ref{fig:compvalid}.

\begin{table*}
	\centering
	\caption{Algorithm configurations investigated in our parametric study \label{TAB-PSDETAILS}}
	\begin{tabular}{ll}
		\hline
		Input & Used parameters\\
		\hline
		$P$ & 10, 20, 50, 100, 200, 500, 1000, 2000, 5000\\
		$S$ & 100, 200, 500, 1000, 2000$^*$\\
		$I$ & 2, 4, 8, 16, 32\\
		$(\max T_{red}-\min T_{red})/S$ & 2, 5, 10, 20, 50, 100\\
		$E$ & $E_{SQD}$, $E_{MAPE}$, $E_{T_{Adj}-SQD}$\\
		\hline
	\end{tabular}
\end{table*}
\subsubsection{Relation between past and future errors}
 
 To investigate our assumption on the relation of past and future error values, we have analysed the collected values in the error cache. In Figure~\ref{fig:exnordu}, we exemplify how the simulated past and future error values (using the $E_{SQD}$ function) vary within a subset of the past/future error cache (which we collected in the previous phase of our evaluation). Here, each dot represents a single simulation run, while the error values were calculated after the twelfth parallel section -- $k=12$, see Figure~\ref{FIG-WF-DESC}. To reduce the clutter in the figure, we present the simulation results using only a subset of the trace fragments that we have identified in the Nordugrid GWA trace ($T^{GWA}_{nordu}\subset T$). The items in the subset were selected so every 50$^{th}$  Nordugrid related trace fragment is shown in the chart: $\{t_i, t_{i+1}\}\in T^{GWA}_{nordu}\to\{t_j,t_{j+50}\}\in T$. Out of the selected fragments, only those are shown in the figure which resulted in relatively low error values. This allows us to better observe the relationship between past and future errors when both error values are low. For example, in case of Figure~\ref{fig:exnordu}, we have used the low error value limit of $10^7$.  Note, the range of this error function for all the simulated trace fragments was: $\forall t in T:1.5\cdot 10^6 < E_{SQD}(W,12,t)< 4 \cdot 10^7$. 
 
 Based on the error cache, in this subsection, we investigated two assumptions that are both important for the algorithm's success: $(i)$ low past error values \emph{likely} pair up with low future error values; and $(ii)$ show that as we approach the end of the workflow, decreasing past error values would more likely pair up with decreasing future error values (ie., as past error values converge towards the final worklfow execution error, future error values also approach a stable hypothetical final value). As a first step, we limited our analysis to fragments with past/future error values below a chosen  \emph{low error value} -- $\tau$ -- threshold:
 \begin{equation}
 T_{filt}^{exp}:=\{t\in T:E(W,t,k)<\tau\lor F(W,t,k)<\tau\}
 \end{equation}
The choice for $\tau$ could ensure that this filtered set contains the trace fragments most likely to be found by Algorithm~\ref{ALG-GFBP} lines 1-\ref{ALG-LINTMIN}. For the error function $E_{SQD}$, we have identified the low error value limit as: $\tau:=2\cdot10^6$.
 
 For our first assumption, we evaluated the likelihood that consecutive fragments with small past error values $E(W,t,k)$ lead to small $F(W,t,k)$ values. We consider two trace fragments ($t_a, t_b\in T: (t_a<t_b)$) consecutive when there are no other trace fragments with starting timestamps in between the starting timestamps of the consecutive ones: $\nexists t_c\in T:(t_c>t_a\land t_c<t_b)$. First, we prepared the subsets of $T_{filt}^{exp}$, that hold more than 80 consecutive timestamps of the trace. The number of consecutive timestamps is calculated as $\mathfrak{T}/\mathfrak{t_{sim}}$ while assuming $\mathfrak{T}$ to be a minute to minimise the impact of the simulation on the complete prediction operation and its users (eg., a workflow enactor) from the autonomous control loop. Next, we observed that in these subsets the \emph{likelihood} of having both minimal future and past error values was 65-86\%. Finally, we also observed that the selection of a lower $\tau$ value could notably decrease the simultaneous presence of below-threshold error values (suggesting that a too precise match for the past/future error leads to over fitting). 
 
 For our second assumption, we evaluated the error cache and we observed that the higher $k$ is the more potential the prediction has. I.e., with increasing $k$ the error values tend towards the low error values, while simultaneously their deviation also decrease. Videos  of this behaviour can be seen in footnotes for the Sharcnet\footnote{https://youtu.be/gozmHoCneyU} and AuverGrid\footnote{https://youtu.be/BzdVcAq4ez8} traces. The recordings show how the past and future error functions of $E_{SQD}$ converge towards the optimal values as a result of the increasing number of completed parallel sections of TCG -- $k$.
 
 To conclude,  we have shown that our assumption on using past error values to indicate the tendency of future ones is well supported by our simulated error cache. In other words, if a trace shows similarity (in terms of an error function) to past workloads then the same trace can be used to estimate future workloads. 
 
\subsubsection{Behaviour analysis of the algorithm}
\begin{table*}
	\centering
	\begin{tabular}{rcccc}
		\hline
		Random selection & $MAPE_E(W,t_g)$ & $MAPE_F(W,t_g)$ & $E_{MAPE}^*$ & $F_{MAPE}^*$\\
		\hline 
		Average ($\forall t_g\in T_G$) & 157.874 &	166.166 & 72.243 & 85.893\\
		Median ($\forall t_g\in T_G$)& 49.180 &	67.825 & 45.174 & 47.002\\
		\hline
	\end{tabular}
	\caption{Baseline results of random trace selection\label{TAB:RANDOM}}
\end{table*}
\begin{figure*}[tb]
	\centering
	\subfloat[$E_{SQD}$]{\includegraphics[width=0.33\textwidth]{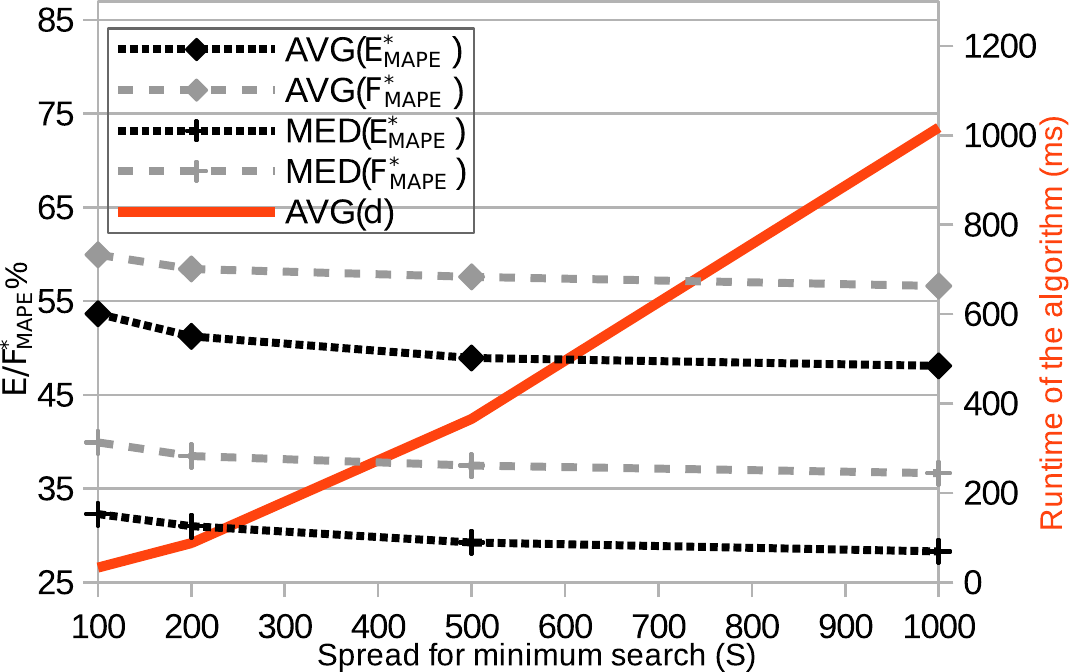}\label{fig:vsS-SQD}}
	\subfloat[$E_{MAPE}$]{\includegraphics[width=0.33\textwidth]{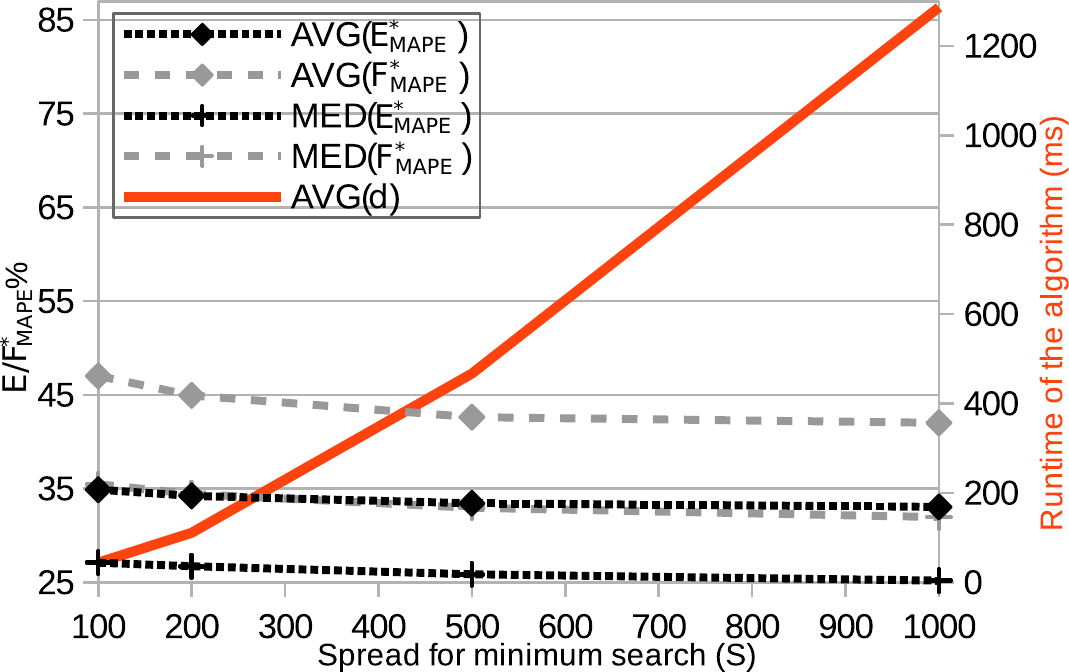}\label{fig:vsSMAPE}}
	\subfloat[$E_{TAlt-SQD}$]{\includegraphics[width=0.33\textwidth]{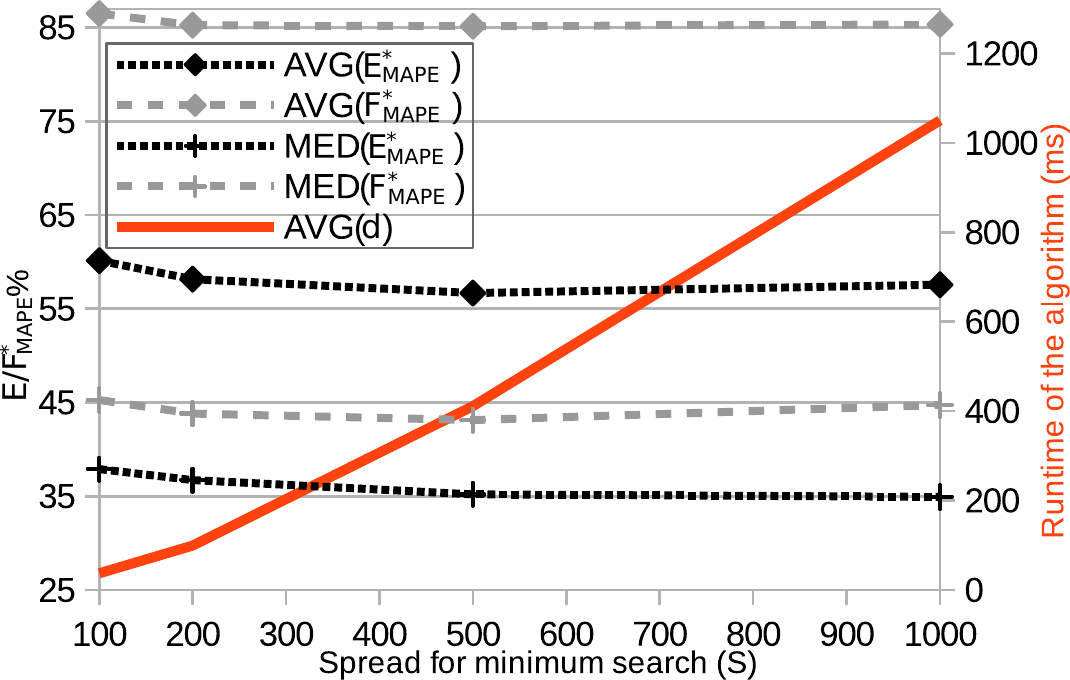}\label{fig:vsSAlt}}
	\caption{The influence of the input $S$ of the algorithm to its performance and accuracy}\label{fig:depfigs-S}
\end{figure*}

\begin{figure*}[tb]
	\centering
	\subfloat[$E_{SQD}$]{\includegraphics[width=0.33\textwidth]{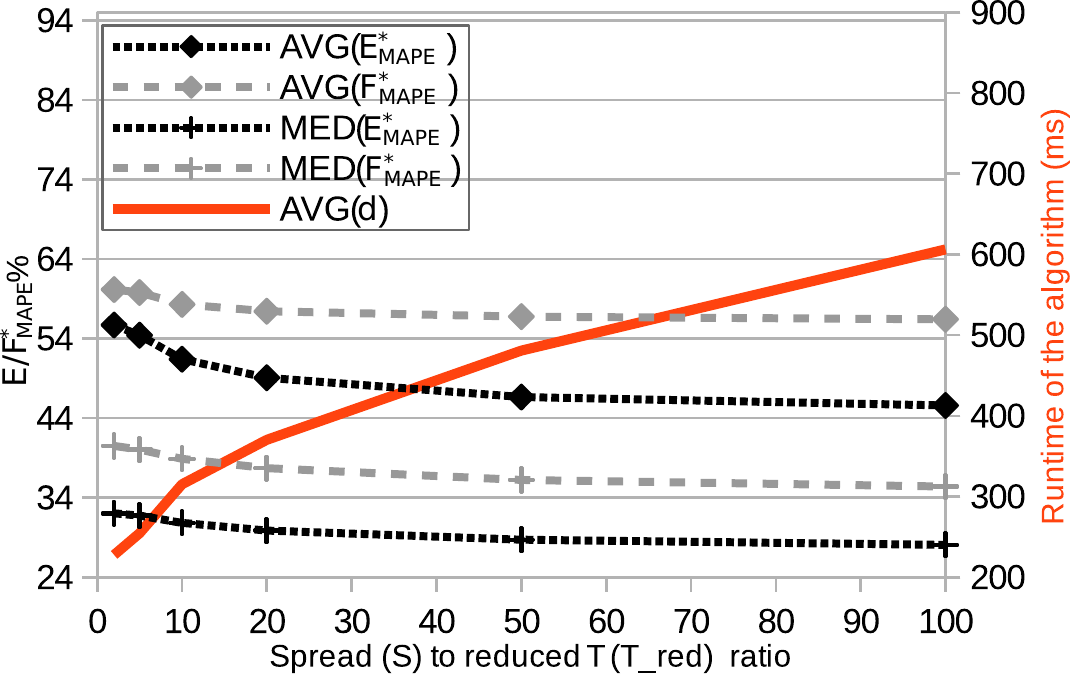}\label{fig:vsWM-SQD}}
	\subfloat[$E_{MAPE}$]{\includegraphics[width=0.33\textwidth]{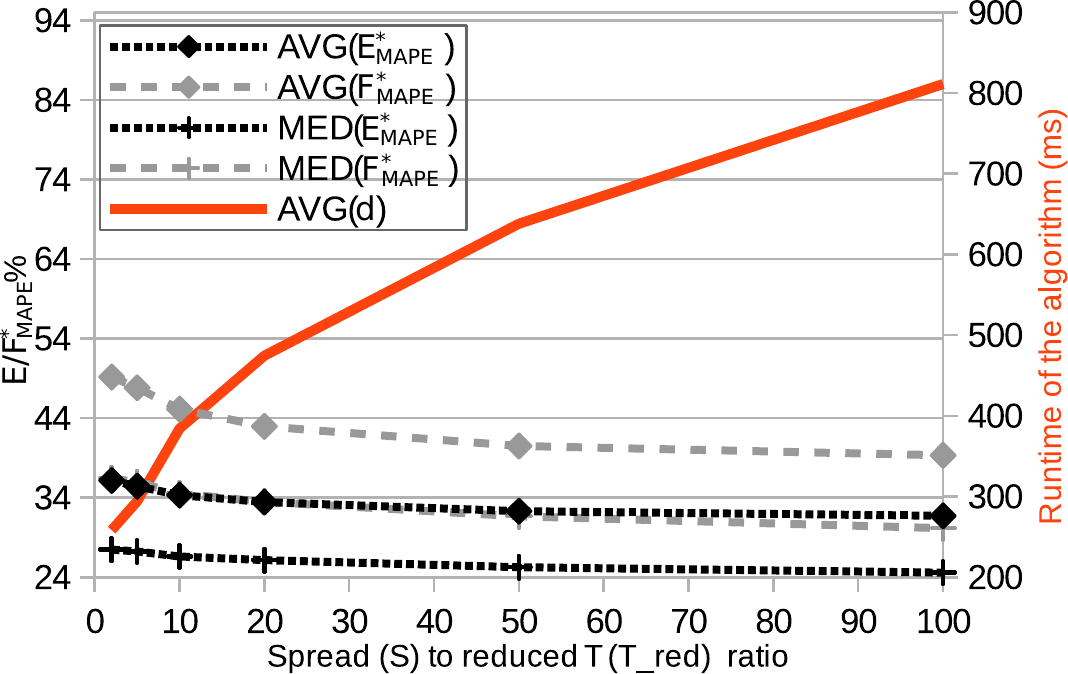}\label{fig:vsWMMAPE}}
	\subfloat[$E_{TAlt-SQD}$]{\includegraphics[width=0.33\textwidth]{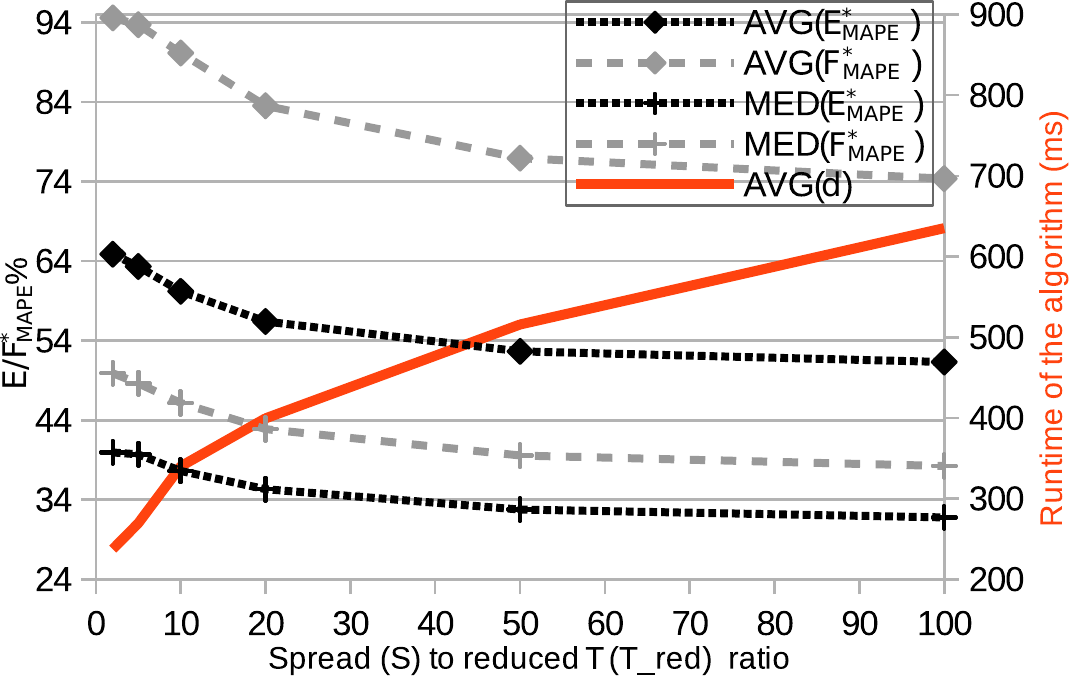}\label{fig:vsWMAlt}}
	\caption{Performance and accuracy impacts of the secondary search window size -- $(\max T_{red}-\min T_{red})/S$}\label{fig:depfigs-WM}
\end{figure*}

In the final phases of our evaluation, we first executed an extensive parametric study on the most important inputs of the algorithm. Table~\ref{TAB-PSDETAILS} shows all the parameters evaluated. In our parametric study, we have ran all input parameter combinations on resources utilizing the private OpenStack cloud of Liverpool John Moores University. We have used 5 VM instances with equivalent configurations as follows: $(i)$ 20GB disks, containing the execution logs and the error cache, $(ii)$ 16GB of memory to allow efficient handling of the cache, and $(iii)$ an Intel Core i7 4790 processor. For each parameter combination, these VMs ran 500 approximations using our previously collected simulated job execution log database (this is depicted in Figure~\ref{fig:compvalid} as the ``\emph{Algorithm behaviour data collection}'' phase). 

For each approximation run, we have randomly selected an approximation target: a golden fragment, which we denote as $t_g\in T_G$, where $T_G\subset T^{GWA}$ denotes the set of all randomly chosen golden fragments. We have retrieved the execution logs related to the golden fragment (ie., the $r_{ob}(j,t_g)$). Then, we applied our algorithm to identify an approximate GWA trace fragment for this particular (golden) TCG workflow instance, given that the instance has progressed to job $j_k$ (ie., the algorithm did not receive the complete execution log, just the observed execution time values that occurred before job $j_k$: $r_{ob}(j_i,t_g):1\leq i<k$). The approximation resulted in a $t_{target}\in T^{GWA}$ trace fragment start time from one of the GWA traces. Finally, we have analysed the relation of $t_{target}$ and $t_g$, as well as the execution time -- $d$ -- of the approximation algorithm under the particular input parameter conditions.

To analyse the capabilities of the algorithm in terms of workload approximation, we first defined the metrics to quantify the accuracy. An ideal solution would be finding exactly the golden fragment however, this is neither feasible nor necessary (as real life traces for the background workload would not be comparable). The aim is to see the degree of similarity between the golden fragment and its approximation (identified by $t_{target}$). We have defined two fundamental metrics for the evaluation of our algorithm:
\begin{description}
	\item[\emph{Execution time level metric:}] First, we used the $E_{MAPE}$ function from Eq.~\ref{EQ-MAPE}. We wanted $E_{MAPE}$ to show the average error (in percentage) between $t_g$ and $t_{target}$, thus during this evaluation, we assigned $r_{ex}(j_i)\gets r_{ob}(j_i,t_g)$ and $r_{ob}(j_i,t)\gets r_{ob}(j_i,t_{target}) $. This allowed us to see the execution time differences the algorithm's predicted $t_{target}$ trace fragment causes in contrast to the golden's. We will denote this special use of the error function as $E_{MAPE}^*$.
	\item[\emph{Error level metric:}] We also compared how do the golden and the approximated trace fragments relate to the real life execution expectations of TCG -- these are the $r_{ex}(j_i)$ values we have identified in the LPDS cloud according to Section~\ref{sec:TCGONLPDS}. For this metric, we again use the mean absolute percentage error method, but this time to see how the error for $t_g$ is approximated by the error of $t_{target}$ at every $k$ value.       
\end{description}
Thus our second, error level, metrics are defined as:
\begin{equation}
MAPE_E(W,t_g):=\sum_{1\le i \le N}\frac{|E(W,t_{g},i)-E(W,t_{target},i)|}{\frac{N}{100}E(W,t_{g},i)}
\end{equation}
for past errors, and
\begin{equation}
MAPE_F(W,t_g):=\sum_{2 \le i \le N-1}\frac{|F(W,t_{g},i)-F(W,t_{target},i)|}{\frac{N-2}{100}F(W,t_{g},i)}
\end{equation}
for future errors. Note: when evaluating our above metrics, we used the same error function as the algorithm -- e.g., $E(W,t_g,i)$ could be in both cases $E_{SQD}(W,t_g,i)$. 

As we have done 500 approximations for each input parameter combination,  we have calculated their overall average and median values (for the following metrics: $E^*_{MAPE}, F^*_{MAPE}, MAPE_E, MAPE_F, d$) to represent the accuracy and performance of the approximation with a particular input set. To put these aggregated values into context, we have evaluated them for a random trace selection approach as well (this served as a baseline for comparing the effectiveness of our technique). Here, for each member of the golden set ($T_G$), an arbitrary trace fragment from the whole trace database was selected as the member's approximation: $t_{target}^{RANDOM}\in T^{GWA}$. The calculated accuracy metrics of this approximation approach are shown in Table \ref{TAB:RANDOM}.

Figures~\ref{fig:depfigs-S} and~\ref{fig:depfigs-WM} show the behaviour induced by the changes in the two most impacting parameters of the algorithm. In total, each figure represents results of over 500 thousand approximations. A plotted point averages the outcomes of all possible parameter configurations except the one that is fixed for the plot. Note, the apparent impact of a parameter is often reduced by averaging (e.g., the best parameter configurations for $S=1000$ lead to $E^*_{MAPE}$ values in the range of 28\%, while the average shown in the figure is over 34\%). Nevertheless, the major trends are still visible also in all cases the approximations of our algorithm yield significantly better results than random selection. Not surprisingly, using the $E^*_{MAPE}$ metric, we have the best results when the algorithm also uses our MAPE based workflow execution time error function ($E_{MAPE}$), while the time adjusted function ($E_{TAlt-SQD}$) performs poorly. The improvements over random selection range between 10-20\% for the median of our execution time level metric. Based on the results, one can also conclude that the secondary search window size (shown in Figure~\ref{fig:depfigs-WM}) is a more important factor (albeit not directly configurable from the algorithm's inputs). Thus, this must be an exposed user configurable parameter in the future.

In general, the $d$ duration of the approximation is negligible (in the range of 31ms-2114ms, with a median of less than 200ms) compared to our assumed 1 minute maximum time for prediction. This leaves enough time for the simulation needs of the algorithm and therefore ensures timely predictions. We have also observed that increasing any of the parameters obviously introduces more calculations, however they do not increase the accuracy in a uniform way. Increasing parameter $I$, the number of iterations increases accuracy in a minimal way. After investigation, we have concluded that the algorithm's exit condition (see line~\ref{ALG-EXITCOND} of Algorithm~\ref{ALG-GFBP}) is often fulfilled by its sub-condition on $\Pi$ -- which was a set constant in all algorithm executions. Increasing the number ($P$) of evaluations for $\phi(x)$ often leads to slightly decreasing accuracy because a higher number of evaluations lead to local minimums.

\begin{figure*}[!t]
	\centering
	\subfloat[$E_{SQD}$]{\includegraphics[width=0.309\textwidth]{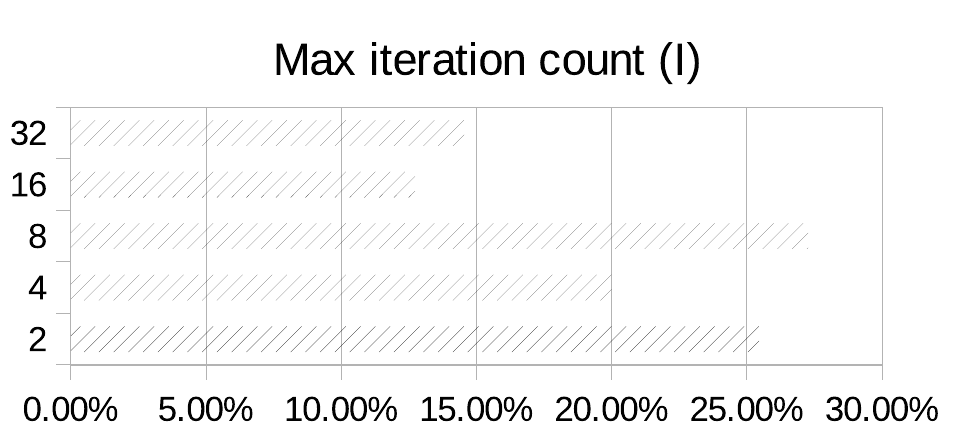}\label{fig:I-SQD}}
	\subfloat[$E_{MAPE}$]{\includegraphics[width=0.31\textwidth]{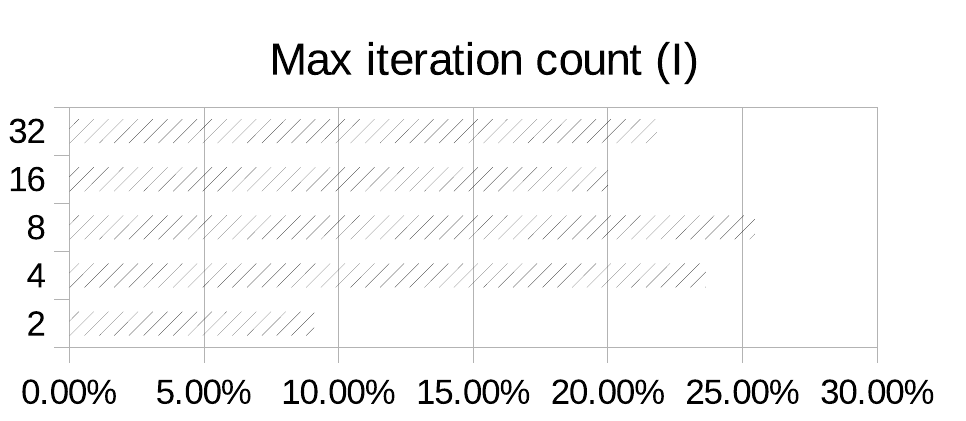}\label{fig:I-MAPE}}
	\subfloat[$E_{TAlt-SQD}$]{\includegraphics[width=0.308\textwidth]{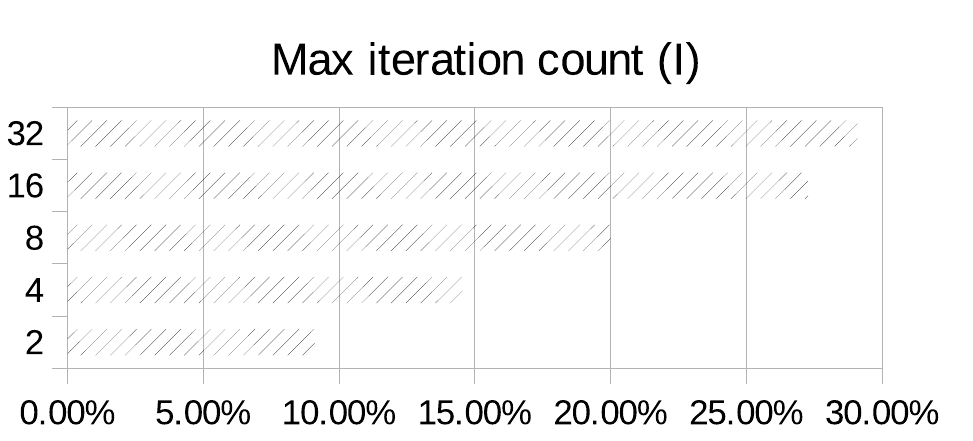}\label{fig:I-TAlt}}
	\caption{The likelihood that the algorithm will produce predictions in the top 5\% depending on input $I$}\label{fig:I-figs}
\end{figure*}
\begin{figure*}[!t]
	\centering
	\subfloat[$E_{SQD}$]{\includegraphics[width=0.31\textwidth]{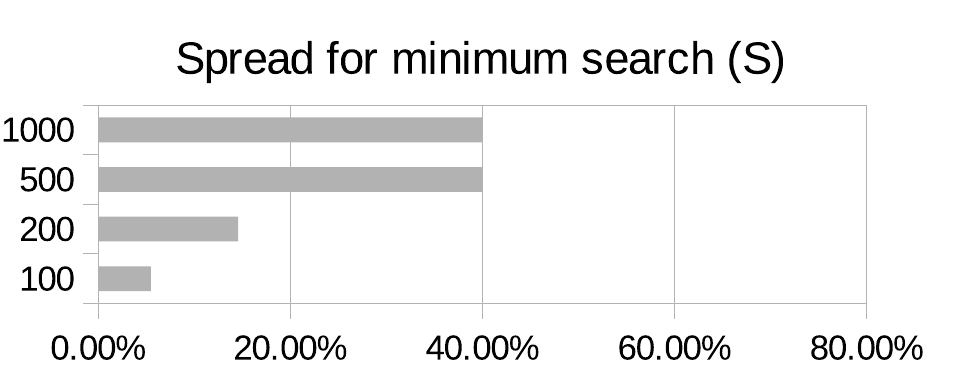}\label{fig:S-SQD}}
	\subfloat[$E_{MAPE}$]{\includegraphics[width=0.302\textwidth]{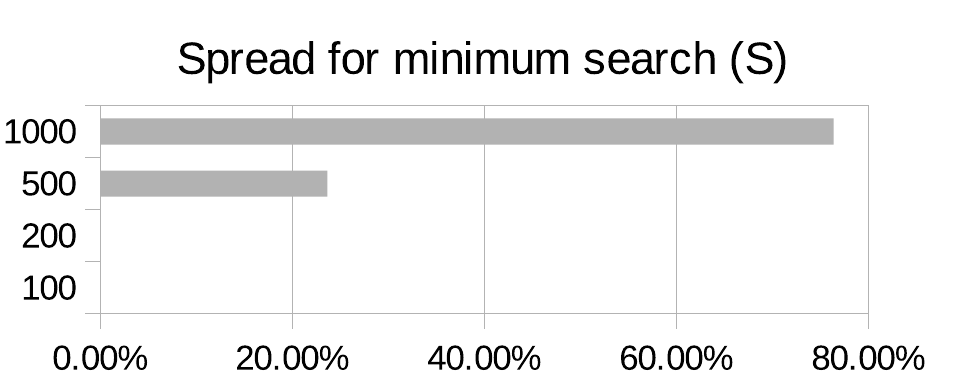}\label{fig:S-MAPE}}
	\subfloat[$E_{TAlt-SQD}$]{\includegraphics[width=0.308\textwidth]{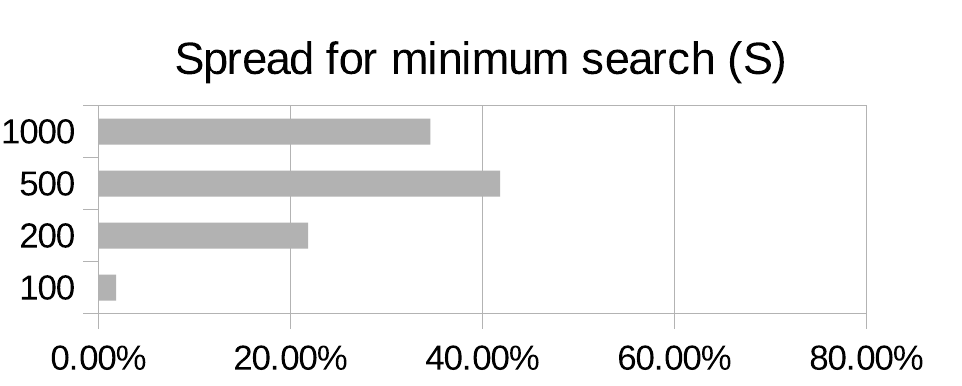}\label{fig:S-TAlt}}
	\caption{The likelihood that the algorithm will produce predictions in the top 5\% depending on input $S$}\label{fig:S-figs}
\end{figure*}
\begin{figure*}[!t]
	\centering
	\subfloat[$E_{SQD}$]{\includegraphics[width=0.31\textwidth]{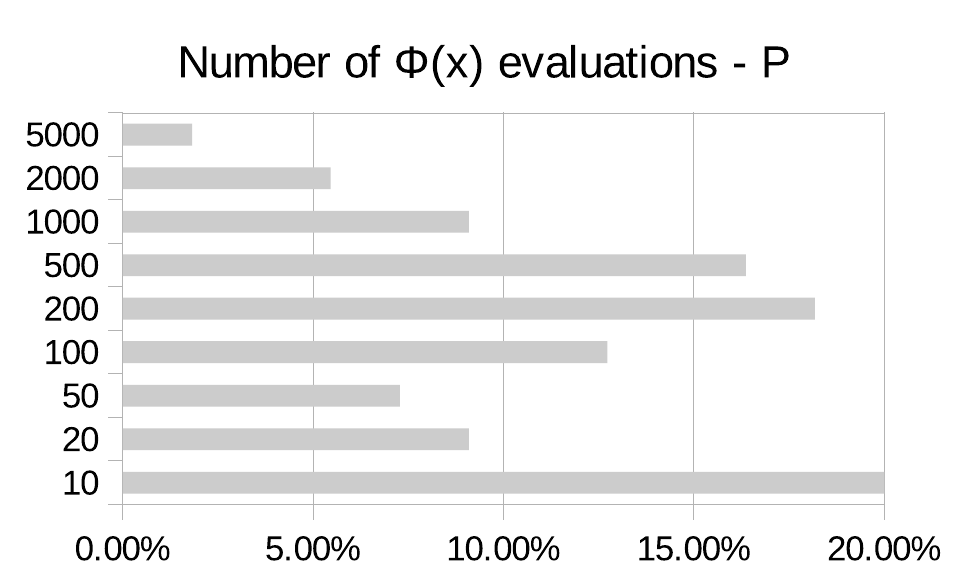}\label{fig:P-SQD}}
	\subfloat[$E_{MAPE}$]{\includegraphics[width=0.308\textwidth]{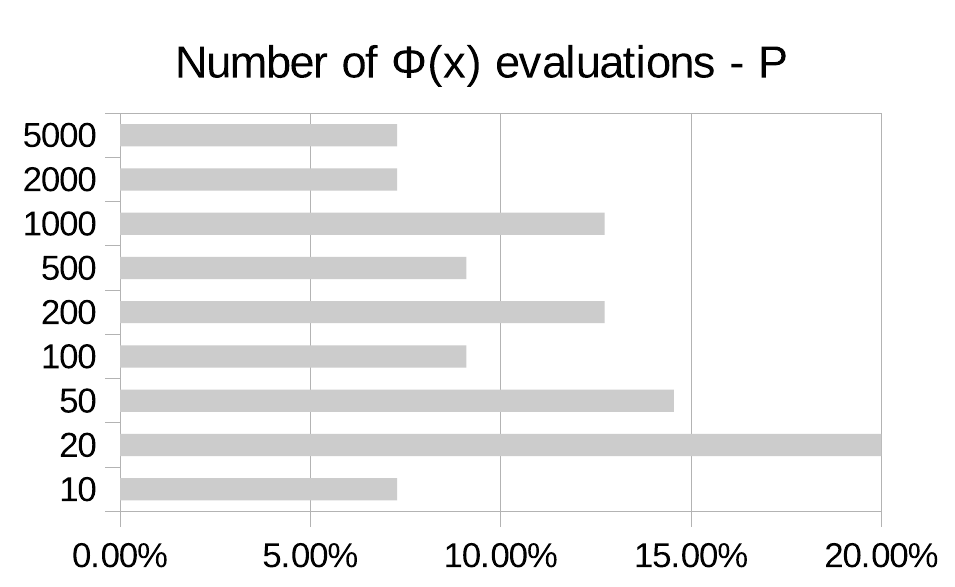}\label{fig:P-MAPE}}
	\subfloat[$E_{TAlt-SQD}$]{\includegraphics[width=0.31\textwidth]{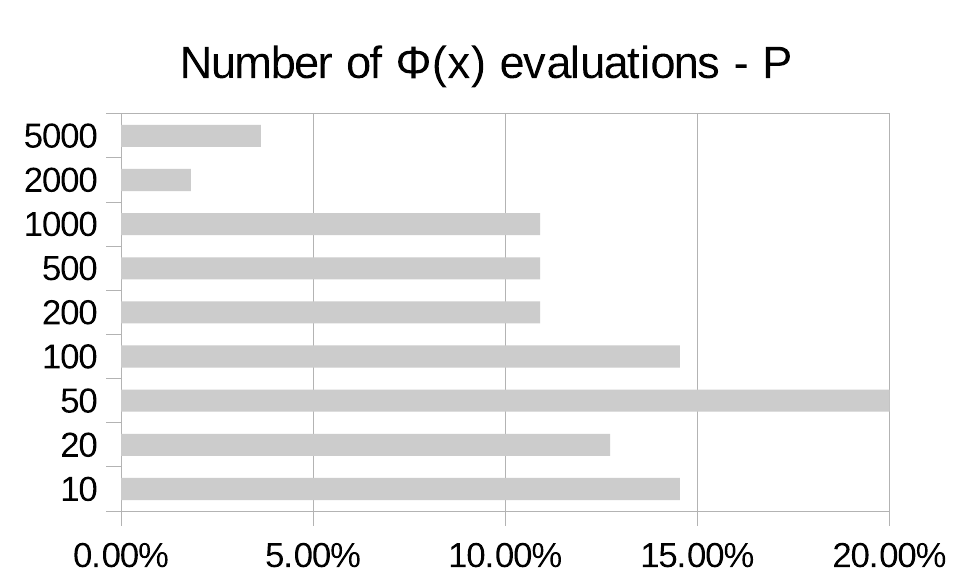}\label{fig:P-TAlt}}
	\caption{The likelihood that the algorithm will produce predictions in the top 5\% depending on input $P$}\label{fig:P-figs}
\end{figure*}
\begin{figure*}[!t]
	\centering
	\subfloat[$E_{SQD}$]{\includegraphics[width=0.31\textwidth]{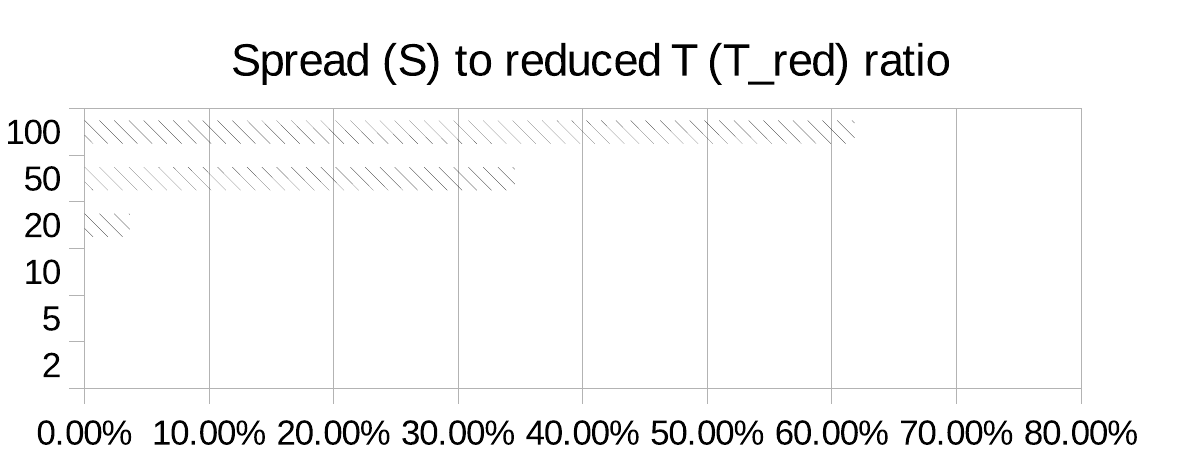}\label{fig:WM-SQD}}
	\subfloat[$E_{MAPE}$]{\includegraphics[width=0.308\textwidth]{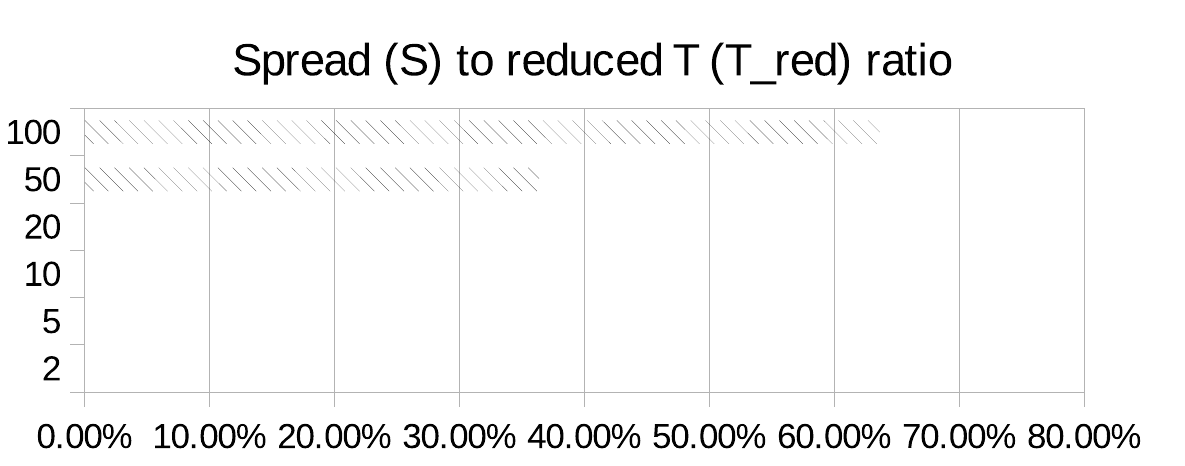}\label{fig:WM-MAPE}}
	\subfloat[$E_{TAlt-SQD}$]{\includegraphics[width=0.31\textwidth]{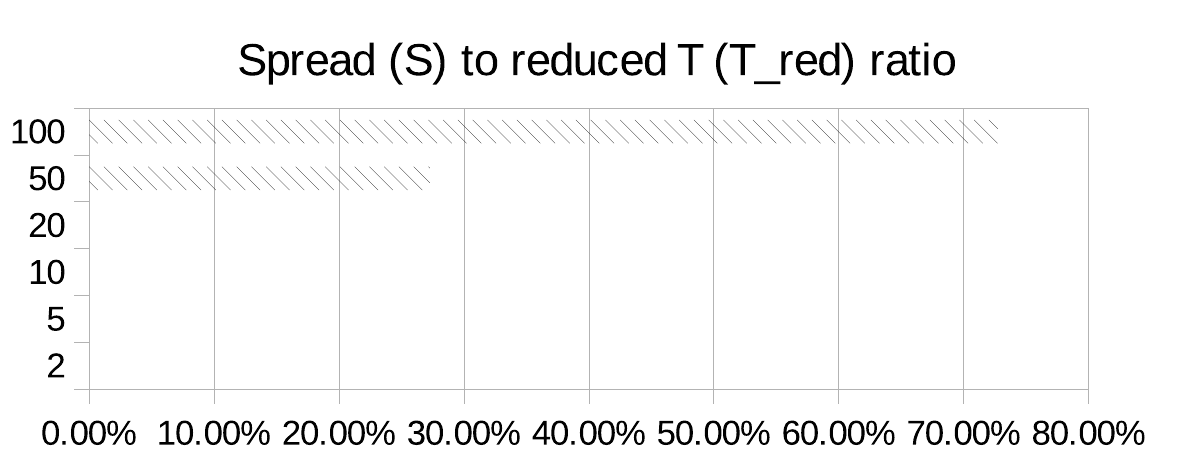}\label{fig:WM-TAlt}}
	\caption{The likelihood that the algorithm will produce predictions in the top 5\% depending on secondary search window size -- $(\max T_{red}-\min T_{red})/S$}\label{fig:T-figs}
\end{figure*}

The effects on accuracy are not conclusive, because the nature of cross-parameter averages shown in Figures~\ref{fig:depfigs-S} and~\ref{fig:depfigs-WM}. To better understand the effects of particular parameters, we have filtered the top 5\% most accurate parameter configurations. Figures \ref{fig:I-figs}, \ref{fig:S-figs} \ref{fig:P-figs} and \ref{fig:T-figs} show the likelihood for a particular parameter value to be represented in the filtered list. The figures also analyse how the particular error functions benefit from the other input parameters of the algorithm. Note, the average $d$ duration of the algorithm's evaluation  with the parameter combinations from the top 5\% was 2333ms.

These figures suggest, that selecting an $I$ value between 8-32 could bring slight benefits in accuracy. With regards to $S$, we have seen that values over 500 are beneficial. Our parametric study evaluated for $S=2000$ as well, but this increased $S$ value did not bring significant enough improvements to compensate for the additional time spent on evaluating the algorithm (i.e., the $d$ increased threefold for an average accuracy increase of less than 1\%). Next, we have analysed the effect of the number of $\phi(x)$ evaluations ($P$). As with our previous experiments, $P$ have had an inconclusive effect on accuracy (which highlights that the algorithm would greatly benefit from techniques that avoid the traps of local minimums). In general, lower $P<500$ values proved more accurate, especially values of 20-50 were strongly represented in the top 5\%. Finally, we have concluded that the secondary search window size has the biggest impact on accuracy, which is to be set between 50-100 times the size of the primary search window $S$. We recommend the following values: 
\begin{eqnarray}
	P&=&20\nonumber\\
	I&=&32\nonumber\\
	S&=&1000\nonumber\\
	\frac{\max T_{red}-\min T_{red}}{S}&=&50
\end{eqnarray}

\begin{table*}
	\centering
	\begin{tabular}{lcc}
		\hline
		Error function	& \multicolumn{2}{c}{Coefficient of determination}\\
		\multicolumn{1}{c}{$\downarrow$} & $R^2(E(t_g),E(t_{target}))$ & $R^2(F(t_g),F(t_{target}))$\\
		\hline 
		$E_{SQD}$ & 0.824 &	0.176\\
		$E_{MAPE}$ & 0.656 & 0.015\\
		$E_{TAlt-SQD}$ & 0.696 & 0.267\\
		\hline
	\end{tabular}
	\caption{Coefficient of determination ($R^2$) for the recommended input combination with the various error functions \label{TAB:RSQ}}
\end{table*}

\begin{figure*}[!t]
	\centering
	\subfloat[Past errors]{\includegraphics[width=0.389\textwidth]{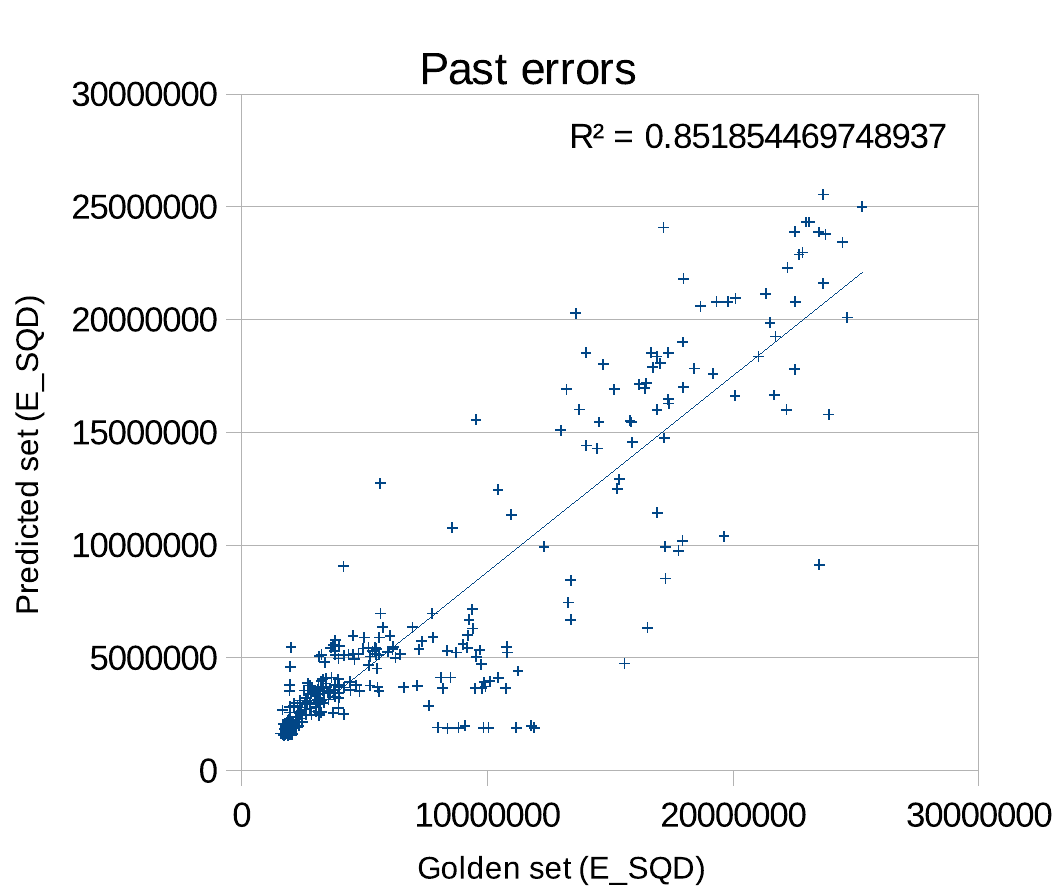}\label{fig:SQD-P-CORR}}
	\subfloat[Future errors]{\includegraphics[width=0.405\textwidth]{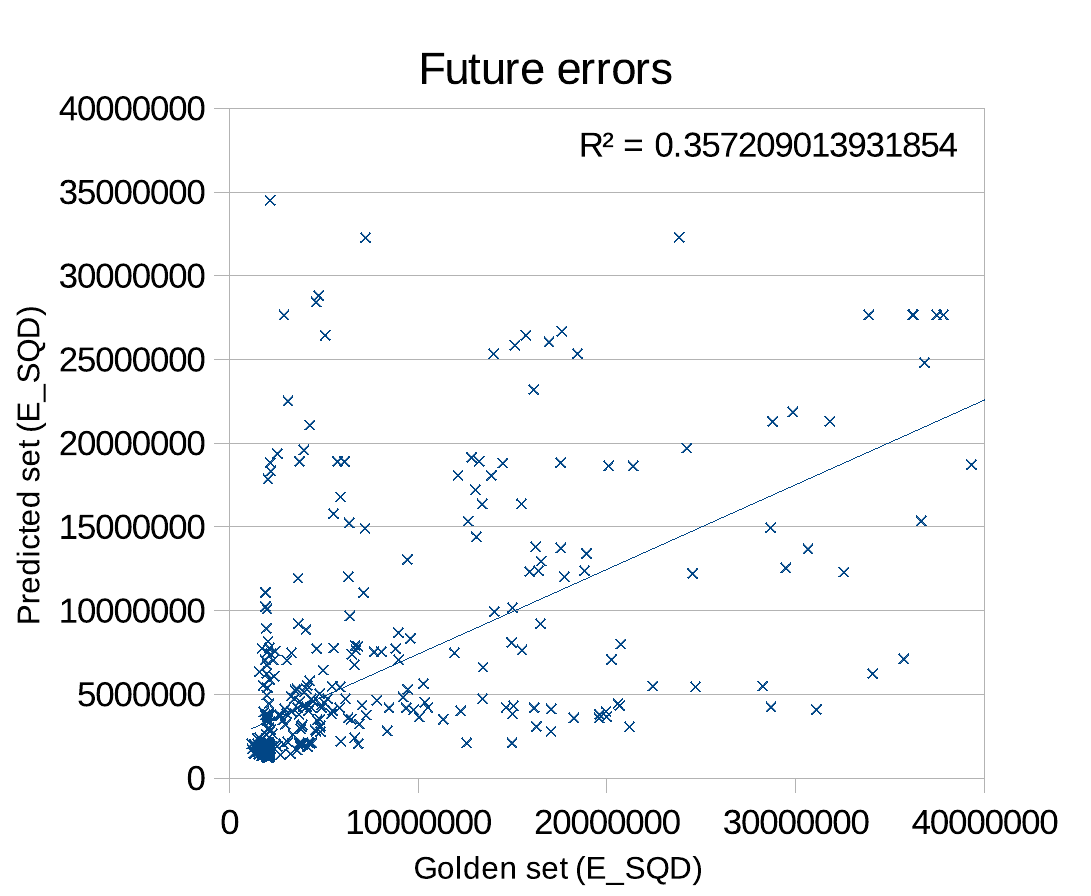}\label{fig:SQD-F-CORR}}
	\caption{Correllation analysis of the $E_{SQD}$ values for the corresponding items from the golden and predicted sets}\label{fig:CORR-SQD}
\end{figure*}

Based on the recommended input values, we have analysed how the past and future errors for each 500 $t_g\in T_G$ correlated with the error values acquired for the corresponding $t_{target}$ predicted fragments. To evaluate the level of correlation, we used the statistical indicator called coefficient of determination: $R^2$. The results are shown in table~\ref{TAB:RSQ}. The algorithm finds strongly correlating approximations for past errors, while future errors show weaker correlation patterns. The weakest correlation was observed for the $E_{MAPE}$ error function, which shows that the function is too focused on the particular $r_{ob}$ values. On the other hand, the error function of $E_{TAlt-SQD}$ leads to the best future error predictions, we assume this performance is likely caused by the function's stronger reliance on job order (and indirectly time). The best $R^2$ results we have obtained are presented in Figure~\ref{fig:CORR-SQD}. These were using slightly different input parameter values (namely $S=500, P=50$) than we recommended based on our statistical evaluation. Amongst our future work, we plan to investigate techniques that would allow the algorithm to auto tune its parameters to get better correlating past and future error predictions for particular workflows.

\section{Conclusions}\label{sec:conclusion}

IaaS clouds hide the complexity of maintaining the physical infrastructure, but there are many application areas that need additional knowledge of the underlying cloud systems in order to support their activities. Workflow enactment is one of these areas that could benefit from detecting behavioural changes in the underlying system. Therefore, this article aimed at studying performance issues related to the background load and proposes a methodology for its estimation. 

We followed the concept of a load prediction method based on the combination of historic traces to improve execution quality. We proposed an algorithm for realising the load prediction at runtime so that performance constraints are observed. We proposed these predictions to select more suitable execution environments for scientific workflows, hence we evaluated this approach using a biochemical application with a state of the art simulator using historic traces from a widely used archive. We have shown that our assumption of using past error values to indicate the tendency of future ones is partially supported by our simulations. Thus, if a trace shows similarity to past workloads then the continuation of the same trace has a potential to be used as an estimate for future workloads.

In our future work, we aim at analysing further areas to employ our algorithm (e.g., to support cloud orchestration, brokering). We also aim at refining our algorithm through multiple approaches: $(i)$ revise the method to select fragments with likely better future matches, $(ii) $ analyse the impact of other error functions, $(iii)$  explore, if certain simulators are better suited for modelling particular clouds and offering better support to our prediction algorithms. Finally, we also plan to enable other (non-workflow-like) long running applications (e.g., commercial web traffic) to offer their inputs to our prediction technique. This direction would allow us to consider broadening the scope of our predictions from private clouds (that could be easily modelled in feasible time with current simulators) to some commercial clouds as well.

\section*{Acknowledgements}
The research leading to these results was supported by the Hungarian Government and the European Regional Development Fund under the grant number GINOP-2.3.2-15-2016-00037 ("Internet of Living Things"), and it was supported by the Janos Bolyai Research Scholarship of the Hungarian Academy of Sciences.

\bibliographystyle{unsrt}
\bibliography{biblio}

\end{document}